\documentclass[%
 aip,
 amsmath,amssymb,
preprint,10pt,%
]{revtex4-2}

\usepackage{graphicx}
\usepackage{dcolumn}
\usepackage{bm}

\usepackage[utf8]{inputenc}
\usepackage[T1]{fontenc}
\usepackage{mathptmx}
\usepackage{xcolor}

\usepackage[colorlinks=true,citecolor=blue,linkcolor=blue,urlcolor=blue]{hyperref}

 \newcommand{\ie}{i.e.}
 \newcommand{\eg}{e.g.}

 \usepackage{color}

\begin{document}


\title{Effect of pulse width on the dynamics of a deflated vesicle in unipolar and bipolar pulsed electric fields} 



\author{Sudip Das}
\affiliation{Aix Marseille Univ, CNRS, Centrale Marseille, M2P2, Marseille, France}
\author{Marc Jaeger}
\affiliation{Aix Marseille Univ, CNRS, Centrale Marseille, M2P2, Marseille, France}
\author{Marc Leonetti}
\affiliation{Univ. Grenoble Alpes, CNRS, Grenoble INP, LRP, Grenoble, France}
\author{Rochish M. Thaokar}
\email[]{rochish@che.iitb.ac.in}
\affiliation{Department of Chemical Engineering, Indian Institute of Technology Bombay, Mumbai, India}
\author{Paul G. Chen}
\email[Author to whom correspondence should be addressed: ]{gang.chen@univ-amu.fr}
\affiliation{Aix Marseille Univ, CNRS, Centrale Marseille, M2P2, Marseille, France}


\date{\today}

\begin{abstract}
Giant unilamellar vesicles subjected to  pulsed direct-current (pulsed-DC) fields are promising biomimetic systems to investigate the electroporation of cells. In strong electric fields, vesicles undergo significant deformation, which strongly alters the transmembrane potential, consequently the electroporation. Previous theoretical studies investigated the electrodeformation of vesicles in DC fields (which are not pulsed). In this work, we computationally investigate the deformation of a deflated vesicle under unipolar, bipolar, and two-step unipolar pulses and show sensitive dependence of intermediate shapes on type of pulse and the pulse width. Starting with the stress-free initial shape of a  deflated vesicle, which is similar to a prolate spheroid, the analysis is presented for the cases with higher and lower conductivities of the inner fluid medium relative to the outer fluid medium. For the ratio of inner to outer fluid conductivity, $\sigma_\mathrm{r}$ = 10, the shape always remains prolate, including when the field is turned off. For $\sigma_\mathrm{r} = 0.1$, several complex dynamics are observed, such as the prolate-to-oblate (PO), prolate-to-oblate-to-prolate (POP) shape transitions in time depending upon the strength of the field and the pulse properties. In this case, on turning off the field, a metastable oblate equilibrium shape is seen, that seems to be a characteristics of a deflated vesicle leading to POPO transitions. When a two-step unipolar pulse (a combination of a strong and a weak subpulse) is applied, a vesicle can reach an oblate or a prolate final shape depending upon the relative durations of the two subpulses. This study suggests that the transmembrane potential can be regulated using a bipolar pulsed-DC field. It also shows that the shapes admitted in the dynamics of a vesicle depends upon whether the pulse is unipolar or bipolar. Parameters are suggested wherein, the simulation results can be demonstrated in experiments.
 
\end{abstract}


\maketitle 

\section{Introduction}\label{intro}

Liposomes or giant unilamellar vesicles (GUVs) with bilayer membranes made of phospholipid molecules are considered to be the most popular model for biological cells, especially for red blood cells.~\citep{biros17,bagchi18,imai11,petch21} This is essentially because of the sizes and the membrane properties of these biomimetic objects are similar to those of the biological cells.~\citep{vlahovska2009vesicles}

The electric field has long been acknowledged as an effective external agent to manipulate cells due to its minimally invasive nature and the ease with which the strength and the effect can be altered in a spatiotemporal sense (using a variety of electrode designs and waveforms). Although electrodeformation of vesicles in a uniform electric field is a promising new method for estimating their electromechanical properties,~\citep{dimova15,salipante12} the most significant application of subjecting cells and vesicles to electric fields is electroporation.~\citep{kotnik01,riske05,sadik11,shamoon19}  Pulsed-direct-current (DC) are typically used in cell electroporation for gene and drug delivery, electrochemotherapy, tissue ablation,~\cite{dimova07} and cell fusion for hybridoma studies. The electric field distribution, the transmembrane potential (TMP), and the membrane tension can significantly vary because of deformation. Thus, a fundamental understanding of the effect of DC-pulses on simultaneous deformation and electroporation merits attention.~\citep{dimova07}

Several experimental,~\citep{karin06,dimova09,dimova10,salipante12,salipante14} analytical,~\citep{lin13,dimova15,salipante14} and computational~\citep{mcconnell15sm,mcconnell15,mcconnell13,Hu16,Veerapaneni16,ebrahim15,ebrahim15a} studies suggested that a vesicle, when suddenly subjected to a DC electric field, undergoes deformation. This electrodeformation of the vesicle depends upon several factors. The most important dimensionless factors of experimental relevance are the non-dimensional strength of the field, $\beta$ (the ratio of electric to viscous force), and the ratio of inner to outer fluid conductivity, $\sigma_\mathrm{r}$. For $\sigma_\mathrm{r}$ = 10, irrespective of the field strength $\beta$, the shape always remains prolate.~\citep{karin06} For $\sigma_\mathrm{r} = 0.1$ and $\beta < \beta_c$, with an initially prolate-shaped vesicle, a small decrease in aspect ratio is observed at short time scales, whereafter, under continued application of the electric field,  the aspect ratio of the vesicle increases. In this case, the vesicle shape always remains prolate.~\citep{mcconnell15sm,ebrahim15a} On the other hand, for $\sigma_r=0.1$ and $\beta > \beta_c$, a prolate deflated vesicle undergoes prolate to oblate shape transition at a short time and later oblate to prolate shape transition, which is commonly called the prolate-oblate-prolate transition (in time).~\citep{karin06,mcconnell13,salipante14,ebrahim15,ebrahim15a,mcconnell15sm,Veerapaneni16,Hu16} In this prolate-oblate-prolate transition, a vesicle also exhibits complex shape deformations alongside the oblate deformation, such as disk-like and cylindrical deformations.~\citep{karin06,ebrahim15,ebrahim15a}

The mechanism of formation of these cylindrical shapes became clear only when rigorous simulations were conducted.  Thus,  a few 2D~\citep{mcconnell15sm,mcconnell15,mcconnell13,Hu16} and 3D~\citep{Veerapaneni16,ebrahim15,ebrahim15a} computational analyses could underpin the physics of the problem. The mechanism was elaborated by McConnel {\it{et al.}}~\citep{mcconnell15sm} as well as by Das and Thaokar~\citep{das18jfm,das18} although in the context of elastic capsules. It is now understood that the Maxwell stress, which is compressive at the poles and the equator when the time is on the order of membrane charging time, at a high electric field, strength is responsible for the cylindrical shape (squaring in 2D) of vesicles.

Moreover, these simulations also indicated that for strong fields, the tension in the membrane could be maximum at a location intermediate between the poles and the equator, indicating a possibility of mechanical failure. It is conjectured that although the TMP remains maximum at the poles, the occurrence of high tension at the intermediate locations between the poles and the equator could lead to poration at those locations. This possibly explains the experimental observation of poration seen at the edges of the cylinder in the case of vesicles subjected to strong DC-pulses with the salt added to the medium.~\citep{karin06} The cylindrical (or spindle like)  deformation observed in vesicles with a conductivity of the inner medium greater than the outer medium\cite{karin06} has not been explained by simulations, which do not account for electroporation effects. Thus it appears that for this case, the cylindrical shape formation is assisted by simultaneous electroporation of the membrane.

Through simulation, Salipante and Vlahovska,~\citep{salipante14} although did not analyze in detail, do report that a vesicle which attains a prolate shape at the end of a pulse could be forced into an oblate shape when the pulse is switched off due to the difference in the conductivity of the inner and outer media. Similarly, McConnel {\it{et al.}}~\citep{mcconnell13} report observing oblate shapes in DC fields, at long times, just above the critical electric field required for observing the prolate-oblate transition. It is argued here that the compressive electric stresses at the equator at long times are unable to overcome the bending stresses, whereby the vesicles are ``locked'' into an oblate shape. However, they report that when the symmetry conditions, imposed in the numerical algorithm, are removed, the long time oblate shape for the specific electric fields disappears. In fact, the critical electric field to observe prolate-oblate transition also reduces significantly. It should be mentioned here, though, that these simulations were limited to 2D.

The methods used for the simulations of vesicle electrohydrodynamics are mainly boundary integral method and immersed interface method for single~\citep{Veerapaneni16,mcconnell13,mcconnell15,mcconnell15sm,ebrahim15,morshed18} and multiple vesicles.~\citep{veerapaneni19pair} These simulations have been performed to address strong nonlinearities arising from shape deformations, strong electric field effects, and nonlinear membrane mechanics. \citet{veerapaneni19pair} present a 2D algorithm for electrohydrodynamic simulation of vesicles and observe several interesting results such as pairing, circulatory, and oscillatory motion of a pair of vesicles, that could be ascribed to dipolar interaction. In these pairwise interactions, the tangential flow is generated at the vesicle surface due to tangential stresses on account of charge migration and the extensional flow generated by the vesicle deformation. These pairwise interactions and other aspects of electrodeformation and the electroporation under strong DC electric pulses were recently reviewed by Perrier, Rems, and Boukany.~\citep{boukany17}

In experimental and theoretical analyses,~\citet{lin13} and \citet{dimova15} studied the relaxation of initially stretched vesicles. In their study, the vesicle is first stretched by applying a uniform electric field. Since the internal fluid medium is more conducting, the vesicle takes a prolate shape. The field is then turned off, allowing the vesicle to relax and attain a spherical shape.

Although uniform pulsed-DC electric fields have long been used in electroporation, recently, the use of bipolar pulsed-DC electric fields have become popular.
Bipolar pulses are alternating pulses of opposite polarity, separated either by a definite time period, which could be zero and offer much better control over the temporal evolution of the TMP. When a unipolar pulse is used, there can be undesired poration after the pulse has been switched off. On the other hand, the use of bipolar pulse can lead to a faster discharge of the membrane, thereby effectively controlling the TMP of the membrane.  This is essentially due to the canceling effect of the opposite sign of the second pulse.~\citep{pakhomov14,nath2020development}
From an application point of view, bipolar pulses are known to result in symmetrical distribution of pores and better survivability of cells while achieving the same degree of electroporation as conventional unipolar pulses.~\citep{tekle91,sweeney16,murovec16,kotnik01}

The existing experimental studies are only on quasispherical vesicles (in the entropic regime), which can attain a spherical shape when the stresses on them are removed. Conducting computational analysis considering both the entropic and enthalpic regime requires complex treatment~\cite{seifert99,sinha18} and has not yet been implemented in numerical simulations and is not attempted here as well.

On the contrary, the present work investigates the deformation of deflated, non-spherical vesicles under strong DC-pulses in the elastic limit. Both unipolar and bipolar pulses of different pulse-widths are considered. The earlier studies on computational analysis of the electrodeformation of vesicles considered deflated vesicles.~\cite{mcconnell15sm,mcconnell15,mcconnell13,Hu16,Veerapaneni16,ebrahim15,ebrahim15a} These deflated vesicles can be achieved at an elevated temperature or by osmosis.~\citep{misbah09}  In the same spirit, our analysis of the dynamics of vesicles in the pulsed electric field is done for stretched deflated vesicles in the enthalpic (elastic) regime. Experiments on the dynamics of deflated vesicles, especially in shear~\cite{mader06,degonville19} and extensional~\cite{kumar20} flow fields, indicate critical dependence on the reduced volume. Such experimental studies have not been conducted with pulsed-DC electric fields, and the existing studies are only on quasispherical vesicles.  Thus, our computational analysis  will be relevant if experiments are conducted on deflated vesicles. This analysis will also be relevant for non-spherical biological cells, which are typically deflated and are not quasispherical. 

Biological cells are non-spherical, not as exceptions but more as a rule.~\cite{kakutani93} The red blood cells, for example, are discoidal, while several other biological cells are ellipsoidal. The giant vesicles are increasingly being reconstituted to duplicate biological cells.~\cite{petra21} Biomaterials such as actin forming the interior of a GUV and membrane proteins and active pumps decorating the bilayers of GUVs are examples of attempts in this direction. In this context, understanding the dynamics of deflated vesicles has become increasingly relevant.

In this work, we attempt to address the lacuna in the electrohydrodynamic studies of vesicles, specifically, the effects of the unipolar and bipolar pulsed DC electric fields and the influence of the pulse width. The focus in this work is on the intermediate shapes during the duration of the pulse and the corresponding transmembrane potential. Importantly, the decay of transmembrane potential with corresponding discharging of the membrane charge and shape relaxation after the pulse is switched off is also investigated, aspects which have not been addressed in the literature. 
 
The rest of the paper is organized as follows: Sec.~\ref{model} summarizes the model we use and solution methods. Section~\ref{results} describes detailed simulation results for the shape deformation of a vesicle in a pulsed-DC electric field, with particular attention to  its dynamics in bipolar pulsed-DC electric fields. Concluding remarks are presented in Sec.~\ref{Concl}.

\section{Model formulation and solution methods} \label{model}
\begin{figure}
\begin{center}
\includegraphics[width=1\textwidth]{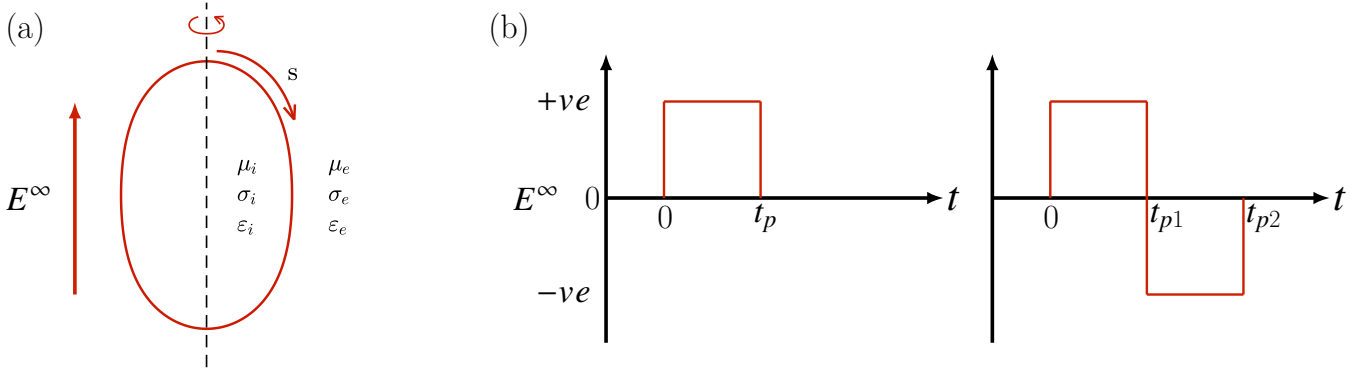}
\caption{(a) Schematic representation of a vesicle in the axisymmetric configuration. Arclength is measured from the north pole and increases in the clockwise direction. The electric field is applied parallel to the axis of symmetry, the upward direction of which is considered to be positive. (b) Schematic representation of the unipolar and bipolar pulsed-DC electric field.}
\label{fig:schematic}
\end{center}
\end{figure}

A deflated vesicle, while suspended in a fluid medium, quickly attains a stress-free configuration. On the sudden application of an external field, the electric field, in this case, the vesicle undergoes deformation. As shown in Fig.~\ref{fig:schematic}(a), there are inner, outer fluid domains ($D$) and the interface ($\Gamma$), and each of them may have different physical properties. The significant parameters are expressed as non-dimensional quantities if possible, such as
\begin{equation}
\lambda = \frac{\mu_\mathrm{i}}{\mu_\mathrm{e}}, \quad \sigma_\mathrm{r} = \frac{\sigma_\mathrm{i}}{\sigma_\mathrm{e}}, \quad \epsilon_\mathrm{r} = \frac{\epsilon_\mathrm{i}}{\epsilon_\mathrm{e}}, 
\end{equation}
where $\lambda$, $\sigma_\mathrm{r}$, and $\epsilon_\mathrm{r}$ are the ratios of viscosities, conductivities, and dielectric constants of the inner (i) and outer (e) fluids, respectively. A lipid bilayer tries to resist the bending of the membrane (m) with the bending modulus $\kappa_\mathrm{b}$. For a lipid bilayer membrane of thickness $\delta_\mathrm{m}$, the effective surface capacitance $C_\mathrm{m}=\epsilon_\mathrm{m}\epsilon_0/\delta_\mathrm{m}$ is the only relevant electrical property of the membrane, as the unporated lipid bilayer has a negligible electrical conductance $G_\mathrm{m} = \sigma_\mathrm{m}/\delta_\mathrm{m}$. In these definitions, $\epsilon_\mathrm{m}$ and $\sigma_\mathrm{m}$ are the membrane dielectric constant and conductance, respectively, and $\epsilon_0$ is the permittivity of free space. Though the lipid bilayer has a typical thickness of $\delta_\mathrm{m}\approx 5$ nm, the popular zero thickness membrane model is adopted in this analysis. 

The unipolar and bipolar pulsed-DC fields are applied to study the dynamics of a deflated vesicle. A schematic representation of unipolar and bipolar pulsed-DC fields is shown in Fig.~\ref{fig:schematic}(b). Upward directing arrows indicate the positive polarity of the electric field, and the opposite direction indicates the reverse polarity. For a constant DC field, the field direction is always positive. If a DC field is applied for a very short duration (10--1000~$\mu$s), the field is termed as the unipolar pulsed-DC field. The duration for which the field remains active is called the pulse width. In Fig.~\ref{fig:schematic}(b), a unipolar pulsed-DC field with pulse-width $t_\mathrm{p}$ is shown. A bipolar pulsed-DC field can be considered as one cycle of a square wave signal, wherein a DC voltage is applied up to a certain time $t_{\mathrm{p1}}$, followed by a change in the sign of the applied DC voltage for time $t_{\mathrm{p1}}<t<t_{\mathrm{p2}}$, whereafter the pulse ends. The amplitude of the electric field remains the same in both parts of the square wave cycle, and we consider $t_{\mathrm{p2}}=2 t_{\mathrm{p1}}$. In experiments, these unipolar and bipolar pulsed-DC fields can be created by a programmable pulse/pattern generator.

Over a typical experimental time scale, the area and volume of a vesicle are preserved; therefore, an important parameter to set is the excess area $\Delta$, which defines the deformability of the vesicle. The area of a vesicle is
\begin{equation}
    A = (4\pi + \Delta) a^2,
\end{equation}
where $a=(V/\frac{4}{3}\pi)^{1/3}$ is the radius of a sphere of equal volume $V$, which is used as the characteristic length in the analysis. As the volume and surface area of a vesicle are preserved, a spherical vesicle ($\Delta = 0$) will always act as a rigid sphere. To undergo deformation, a vesicle must have a shape deviated from a sphere, \ie, a deflated vesicle. A commonly used equivalent parameter to define a deflated vesicle is the reduced volume
\begin{equation}
    \nu = \frac{3 \sqrt{4\pi} V}{A^{3/2}},
\end{equation}
defined as the ratio of the volume of the vesicle to the volume of a sphere having an equal surface area. The reduced volume ranges between $0$, representing a completely deflated vesicle, and $1$, representing a spherical vesicle.~\cite{kumar20,mader06,degonville19,misbah09} The excess area and the reduced volume are related by
\begin{equation}
    \nu = \left(1 + \frac{\Delta}{4\pi}\right)^{-3/2}. 
\end{equation}
Thus, such an area and volume conserving deflated vesicle can never admit a spherical shape in sharp contrast to quasi-spherical vesicles.

The fluid at the domain inside and outside of a vesicle and at the interface experience creeping flow motion; therefore, the Stokes equations can be used as the model equations to study the dynamics of the vesicle,
\begin{equation}\label{eq:stokes}
 \mathbf{\nabla}\cdot \mathbf{u}_k = 0, \quad \mathbf{\nabla}\cdot \mathbf{T}_k^\mathrm{H} =  \mathbf{0}, \quad k= \mathrm{i,e}, \quad \mathbf{x} \in D,
\end{equation}
where $\mathbf{u}$ is the fluid velocity, $\mathbf{T}_k^\mathrm{H}\equiv - p_k\mathbf{I} +\mu_k \left [   (\mathbf{\nabla}\mathbf{u})_k + (\mathbf{\nabla}\mathbf{u})_k^T \right ]$ the hydrodynamic stress tensor, and $p_k$ the fluid pressure.  In vesicles, the surface-area incompressibility is imposed as divergence-free surface velocity, \ie, 
\begin{equation}\label{eq:areainc}
    \mathbf{\nabla}_\mathrm{s}\cdot \mathbf{u}^\Gamma = 0, \quad \mathbf{x}\in \Gamma, 
\end{equation}
which, in turn, results in interfacial tension and appears as a Lagrange multiplier, $\zeta$, in the membrane mechanics.  The tangential surface gradient is defined as $\mathbf{\nabla}_\mathrm{s} = (\mathbf{I}-\mathbf{n n})\cdot \mathbf{\nabla}$, where $\mathbf{\nabla}$ is the  $\mathbb{R}^3$ gradient and $\mathbf{n}$ is the outward unit normal vector on the interface. The dynamic boundary condition at the interface is associated with the nonlinear fluid-structure interaction, the ambient flow, and the membrane mechanics, \ie,
\begin{equation}\label{eq:bc}
 \mathbf{f}^\mathrm{m} + \Delta\mathbf{f} = 0, \quad \mathbf{x}\in \Gamma,
\end{equation}
where $\mathbf{f}^\mathrm{m}$ is the nonhydrodynamic traction acting at the interface, and $\Delta \mathbf{f} =(\mathbf{T}_\mathrm{e}^\mathrm{H}-\mathbf{T}_\mathrm{i}^\mathrm{H})\cdot \mathbf{n}$ is the net hydrodynamic traction exerted by the fluid on the membrane. In the electrodeformation of a vesicle, the nonhydrodynamic tractions are the membrane's resistance to bending, the tension due to the imposed area incompressibility, and the traction due to the applied electric field. Each of these forces is associated with its own time scale.\citep{mcconnell15sm,Thaokar_2020} The capacitor model is used to solve the electric tractions at the interface.\citep{debruin99,seiwert12,young15} The electric stresses caused by the contrast in electrical properties generate hydrodynamic flow in the surrounding fluid on a time scale $\tilde t_\mathrm{h}=\mu_\mathrm{e}/(\epsilon_\mathrm{e}\epsilon_0E_0^2)$, where $E_0$ is the intensity of the applied electric field. The time scale associate with the membrane bending rigidity $\kappa_\mathrm{b}$ is $\tilde t_\mathrm{b} = \mu_\mathrm{e} a^3/\kappa_\mathrm{b}$. 
 A spherical vesicular capacitor has a time constant $\tilde t_\mathrm{m} = \frac{aC_\mathrm{m}}{\sigma_e} \left [1/2 + \sigma_\mathrm{e}/\sigma_\mathrm{i} \right ]$, also known as capacitor charging time, time taken for $\approx 63\%$ charging of a capacitor.~\citep{schwan92} Under typical experimental conditions, the magnitudes of different time scales are on the order of $\tilde t_\mathrm{h} \approx 10^{-3}$ s and $\tilde t_\mathrm{b}\approx 10$ s.\citep{mcconnell15sm,Thaokar_2020} In this work, we nondimensionalize the time by a measure of the membrane charging time $\tilde t_c=a C_\mathrm{m}/\sigma_\mathrm{e}$, which is on the order of $10^{-3}$ s. This leads to the non-dimensional capacitor charging timescale, $t_\mathrm{m} = \left [1/2 + \sigma_\mathrm{e}/\sigma_\mathrm{i} \right ]$, and correspondingly, the other non-dimensional timescales (\eg, $t_\mathrm{b}=\tilde t_\mathrm{b}/\tilde t_\mathrm{c}$ and $t_\mathrm{h}=\tilde t_\mathrm{h}/\tilde t_\mathrm{c}$) can be obtained appropriately. It should be noted that the bulk fluid charge relaxation time $\epsilon_e \epsilon_0/\sigma_e$ and the Maxwell Wagner relaxation time $\epsilon_o (2 \epsilon_e+\epsilon_i)/(2 \sigma_e+\sigma_i)$ are on the order of $10^{-5}$ s, much smaller than other time scales, and are ignored in the calculations. 

The simulations performed and the results presented are in non-dimensional form. Lengths are scaled with $a$ and time with $\tilde t_\mathrm{c}$. Stresses are nondimensionalized by viscous stress $\tau_\mathrm{c}=\mu_\mathrm{e}/\tilde t_\mathrm{c}$. With these choices, the net nonhydrodynamic traction acting at the interface is given in dimensionless form by
\begin{equation}\label{eq:fm}
 \mathbf{f}^\mathrm{m}  = \chi\left[2\Delta_\mathrm{s}H + 4H(H^2-K)\right]\mathbf{n} - 2H \zeta \mathbf{n} + \mathbf{\nabla}_\mathrm{s}\zeta + \beta\ \mathbf{f}^\mathrm{E},
\end{equation}
where $H$ and $K$ are the mean and Gaussian curvatures, and $\Delta_\mathrm{s} H=\mathbf{\nabla}_\mathrm{s} \cdot \mathbf{\nabla}_\mathrm{s}H$ is the Laplace-Beltrami operator of the mean curvature, which contains the fourth derivative of the surface position. The dimensionless bending rigidity $\chi$ is defined as $\chi = 1/t_\mathrm{b}=C_\mathrm{m}\kappa_\mathrm{b}/(\sigma_\mathrm{e}\mu_\mathrm{e}a^2)$, and the (dimensionless) electric field strength $\beta$ is defined as $\beta = 1/ t_\mathrm{h}=aC_\mathrm{m} \epsilon_\mathrm{e}\epsilon_0E_0^2/(\sigma_\mathrm{e}\mu_\mathrm{e})$, which measures the relative importance of electric stress to viscous stress. One may define an electric capillary number $Ca_\mathrm{E}=\beta/\chi=\epsilon_\mathrm{e}\epsilon_0E_0^2a^3/\kappa_\mathrm{b}$ to characterize the relative strength of the shape-distorting electric force to the shape-preserving bending force.

Accurate computation of membrane bending forces poses serious algorithmic and numerical challenges as it involves numerically evaluating a fourth-order derivative of the surface position.\citep{Guckenberger_CPC_2016} In this work, we make use of the numerical method developed in Refs.~\citenum{boedec11} and \citenum{trozzo15} to compute the bending stress and the membrane tension. Therefore, we outline below only how to calculate the electric traction $\mathbf{f}^\mathrm{E}$. 

In the course of deformation in axisymmetric electric fields, a vesicle remains axisymmetric with the symmetry axis parallel to the electric field lines.  
Therefore, the electrohydrodynamic deformation of a vesicle can be computed using the axisymmetric boundary integral method. 
The inner and outer electric potentials of a vesicle satisfy Laplace's equation $\nabla^2 \phi_\mathrm{i,e}=0$. Using Green's theorem, 
the solution of Laplace's equation at the interface ($\mathbf{x}\in \Gamma$) can be formulated in the integral form,

\begin{subequations}\label{eq:pot}
\begin{align}
 \frac{1}{2}\phi_\mathrm{i}(\mathbf{x}_0) =& \int_s [G^\mathrm{E}(\mathbf{ x},\mathbf{ x}_0)\mathbf{ \nabla}\phi_\mathrm{i}(\mathbf{ x})\cdot \mathbf{ n}(\mathbf{ x})-\phi_\mathrm{i}(\mathbf{ x})\mathbf{ n}(\mathbf{ x})\cdot \mathbf{ \nabla}G^\mathrm{E}(\mathbf{ x},\mathbf{ x}_0)] \mathrm{d}S(\mathbf{ x})\label{eq:pot1},  \\ 
 \frac{1}{2}\phi_\mathrm{e}(\mathbf{ x}_0) =& \phi^\infty (\mathbf{ x}_0)-\int_s [G^\mathrm{E}(\mathbf{ x},\mathbf{ x}_0)\mathbf{ \nabla}\phi_\mathrm{e}(\mathbf{ x})\cdot \mathbf{ n}(\mathbf{ x})-\phi_\mathrm{e}(\mathbf{ x})\mathbf{n}(\mathbf{x})\cdot \mathbf{\nabla}G^\mathrm{E}(\mathbf{x},\mathbf{x}_0)]  \mathrm{d}S(\mathbf{x}) \label{eq:pot2},
\end{align}
\end{subequations}
 where $G^\mathrm{E}(\mathbf{x},\mathbf{x}_0)=\frac{1}{4\pi |\hat{\mathbf{x}}|}$ is the Green function for Laplace's equation, $\phi^{\infty}=-y$ is the applied potential and $\hat{\mathbf{x}}=\mathbf{x}-\mathbf{x}_0$, where $\mathbf{x}$ is the observation or load point and $\mathbf{x}_0$ is the pole or source point.~\citep{mcconnell13,das18jfm} The Ohmic current dominates the electrical current continuity across a lipid bilayer membrane; the non-dimensional form of the electrical current continuity equation is obtained as 
\begin{equation}\label{eq:currrent}
    \sigma_\mathrm{r}E_{n,\mathrm{i}}=E_{n,\mathrm{e}}=\frac{ \mathrm{d} \phi_\mathrm{m}}{\mathrm{d} t}, \quad \mathbf{x}\in \Gamma,
\end{equation}
where, $E_{n,\mathrm{i/e}}=\mathbf{n}_\mathrm{i/e}\cdot \mathbf{E}_\mathrm{i/e}$ is the normal electric field at inner/outer surfaces of the membrane, and $\phi_\mathrm{m}=\phi_\mathrm{i}-\phi_\mathrm{e}$ is the transmembrane potential. The electric stresses at the interface can be obtained through Maxwell's relation $\mathbf{T}^\mathrm{E} =\mathbf{EE}-\mathbf{I}\mathrm{E}^2/2 $. The electric tractions can be calculated as $\mathbf{f}^\mathrm{E} = (\mathbf{T}^\mathrm{E}_\mathrm{e}-\mathbf{T}^\mathrm{E}_\mathrm{i})\cdot \mathbf{n}=\tau_n^\mathrm{E}\mathbf{n} - \tau_t^\mathrm{E}\mathbf{t}$, where
\begin{subequations}\label{eq:estress}
\begin{eqnarray}
\tau_n^\mathrm{E} &=& \frac{1}{2}\left[\left(E_{n,\mathrm{e}}^2-E_{t,\mathrm{e}}^2\right)-\epsilon_\mathrm{r}\left(E_{n,\mathrm{i}}^2-E_{t,\mathrm{i}}^2\right)\right],\\
\tau_t^\mathrm{E} &=& \left[E_{n,\mathrm{e}}E_{t,\mathrm{e}}-\epsilon_\mathrm{r}E_{n,\mathrm{i}}E_{t,\mathrm{i}}\right].
\end{eqnarray}
\end{subequations}

The integral form of the Stokes equations, the boundary integral equation,~\citep{rallison78} gives rise to the fluid velocity at the interface ($\mathbf{x}\in \Gamma$) as
\begin{equation}\label{eq:veleqn}
\begin{split}
\mathbf{u}(\mathbf{x}_0)=\frac{1}{1+\lambda}\frac{1}{4\pi}\int_\Gamma \mathbf{f}^\mathrm{m}(\mathbf{x})\cdot\mathbf{G}(\mathbf{x},\mathbf{x}_0)\mathrm{d}S(\mathbf{x})
+ \frac{1}{4\pi}\frac{1-\lambda}{1+\lambda}\int_\Gamma \mathbf{u}(\mathbf{x})\cdot\mathbf{Q}(\mathbf{x},\mathbf{x}_0)\cdot\mathbf{n}(\mathbf{x})\mathrm{d}S(\mathbf{x}).
\end{split}
\end{equation}
The explicit expressions for the free-space Green's functions for the velocity $\mathbf{G}(\mathbf{x},\mathbf{x}_0)$ and the stress $\mathbf{Q}(\mathbf{x},\mathbf{x}_0)$ are given by
\begin{equation}
 \mathbf{G}(\mathbf{x},\mathbf{x}_0)=\frac{\mathbf{I}}{|\hat{\mathbf{x}}|}+\frac{\hat{\mathbf{x}}\hat{\mathbf{x}}}{{|\hat{\mathbf{x}}|}^3},\quad \mathbf{Q}(\mathbf{x},\mathbf{x}_0)=-6\frac{\hat{\mathbf{x}}\hat{\mathbf{x}}\hat{\mathbf{x}}}{{|\hat{\mathbf{x}}|}^5}.
\end{equation}
The deformation of a vesicle can be obtained by updating the shape with the normal component of the interface velocity. In our computation, we follow the Lagrangian approach in moving the equispaced collocation points representing the interface. Therefore, the interface motion is associated with the interfacial fluid velocity through the kinematic condition
\begin{equation}
\frac{\mathrm{d}\mathbf{x}}{\mathrm{d}t} = \mathbf{u}^\Gamma. 
\end{equation}

In the numerical computation, the electric field solution and the solution of the hydrodynamics are decoupled, owing to the faster electric response time relative to the hydrodynamic response time. Therefore, first, we solve the electric stresses via boundary integral equations [(Eq.~(\ref{eq:pot})] over the shape at the current time. Then we solve the boundary integral equation [Eq.~(\ref{eq:veleqn})] with the calculated electric stresses and other nonhydrodynamic membrane forces. The membrane tension can be formulated as the sum of the tensions arising due to the bending force $\zeta^\mathrm{b}$ and the electric force $\zeta^\mathrm{E}$, and Eq.~(\ref{eq:fm}) can be rewritten as
\begin{equation}\label{eq:fm1}
\begin{split}
 \mathbf{f}^\mathrm{m}  = & \chi \left[2\Delta_\mathrm{s} H + 4H(H^2-K)\right]\mathbf{n} - 2 H\zeta^\mathrm{b} \mathbf{n} + \mathbf{\nabla}_\mathrm{s}\zeta^\mathrm{b} + \beta\ \mathbf{f}^\mathrm{E} - 2 H \zeta^\mathrm{E}\mathbf{n} + \mathbf{\nabla}_\mathrm{s}\zeta^\mathrm{E}=\mathbf{f}^{\mathrm{b}\zeta}+\mathbf{f}^{\mathrm{E}\zeta}.
 \end{split}
\end{equation}
First, the boundary integral equation solving interface velocity [Eq.~(\ref{eq:veleqn})] along with the incompressibility constraint [Eq.~(\ref{eq:areainc})] is solved with $\mathbf{f}^{\mathrm{E}\zeta}$ following explicit algorithm. Then, following the numerical method described in an earlier work\citep{boedec11} for the implicit solution of the interface velocity, we solve the same set of equations with $\mathbf{f}^{\mathrm{b}\zeta}$. The interface velocity due to the $\mathbf{f}^{\mathrm{E}\zeta}$, obtained in the solution of the first part of the calculation, is treated as the $\mathbf{u}^{\infty}$ in the implicit solution of the boundary integral equation, as formulated in Eq.~(16) in Ref.~\citenum{trozzo15}. The severe constraint on time-stepping prohibits the computation with the bending force following the explicit time-stepping algorithm.~\citep{boedec11} Therefore, to carry out computation with a reasonable cost, we follow this two-step approach.

In the computation, $220$ elements and a constant time-step size $\Delta t=10^{-5}$ are considered so that the relative errors in the change in area and volume remain below $0.005\%$ and $0.01\%$, respectively (see Appendix). To maintain the uniform element size, node points are redistributed, and the transmembrane potential, the initial condition for the field solution, is interpolated at the new positions of the node points using cubic spline interpolation. The rigorous validation of the boundary integral codes for the electrostatic~\citep{das18jfm} and hydrodynamic~\citep{trozzo15} parts has been reported in our earlier works.

\section{Results}\label{results}
In typical vesicle electrodeformation experiments, the inner and outer fluids are aqueous; hence, they can be considered to have the same dielectric constant, \ie, $\epsilon_\mathrm{r}=1$. On the other hand, the fluid conductivities (inner and outer) can be conveniently changed by adding electrolytes. In a system with a higher outer fluid medium conductivity than the inner fluid medium $\sigma_\mathrm{r}<1$, vesicles are known to show complex dynamics in uniform electric fields.~\citep{mcconnell13,mcconnell15,mcconnell15sm,Veerapaneni16,ebrahim15,ebrahim15a,veerapaneni19pair,salipante14}

In this work, we explain the underlying mechanism of the deformation in the pulsed-DC electric fields, with a focus on the dynamics in bipolar pulsed-DC electric fields. It should be noted that we consider only the elastic limit, while the entropic membrane fluctuations and associated tension due to thermal undulations are neglected.

Vesicles with an aspect ratio (ratio of major to minor axes) greater than one are termed prolate, and those with an aspect ratio less than one are called oblate shapes. When the aspect ratio of an initially prolate vesicle goes below $1.0$, it is termed the prolate-oblate transition. Similarly, when the aspect ratio changes from less than $1.0$ to greater than $1.0$, it is referred to as the oblate-prolate transition. 

In the dynamics of a vesicle in pulsed DC field, the intermediate shapes can be labeled as A-B-C-D transition in time, wherein A  represents the initial state, always prolate for stress free shape at $\nu =0.9$, B and C are the intermediate shapes, and D is the final equilibrium shape after the pulse is switched off.  Later on, in this A-B-C-D transition, an A to B transition is discussed as A-B- transition, the B to C transition is discussed as -B-C- transition, and the C to D transition is discussed as -C-D transition; mark the position of ``-'' in these notations. For example, the deformation of a prolate vesicle to oblate shape, then prolate, and the final oblate shape is marked as prolate-oblate-prolate-oblate transition in time. In this shape transition the B to C shape change is discussed as -oblate-prolate- transition.

\begin{figure}
\begin{center}
\includegraphics[width=0.5\textwidth]{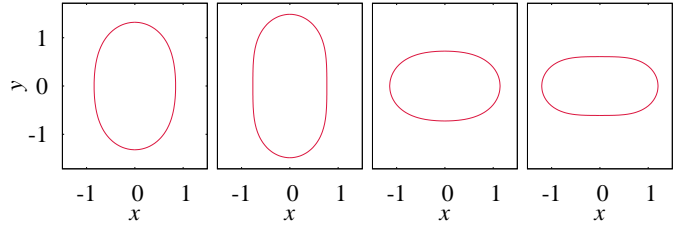}
\caption{Stress-free shapes of the prolate (two shapes at the left) and oblate (two shapes at the right) vesicles with reduced volume $\nu = 0.95$, $0.9$, respectively.}
\label{fig:eqmshapes}
\end{center}
\end{figure}

\subsection{Stress-free shapes}

The Helfrich energy, $E_\mathrm{b}=2\kappa_\mathrm{b}\int H^2\mathrm{d}S$, has been used to study the configurations of deflated vesicles in the absence of any external forces.~\citep{seifert91} Vesicles are known to spontaneously form in an aqueous solution and, at equilibrium, acquire a minimum energy configuration or a stress-free configuration. Depending upon the reduced volume, a vesicle can have a minimum energy prolate ($\nu \ge 0.652$), oblate ($ 0.592 \le \nu \le 0.651$), or stomatocyte ($\nu\le 0.591$) configuration. Our calculations show that corresponding to the reduced volumes $\nu = 0.95$ and $0.9$, the Helfrich energies, scaled with $8\pi\kappa_\mathrm{b}$, for the prolate stress-free shapes are $1.0959$, $1.1913$, respectively, and $1.1068$, $1.2217$, respectively for the oblate stress-free shapes, which are in excellent agreement with the literature values.~\citep{seifert91} These stress-free shapes corresponding to the reduced volumes $\nu = 0.95$ and $0.9$ are shown in Fig.~\ref{fig:eqmshapes}. It should be noted that for a given reduced volume if a vesicle has a starting shape that does not correspond to the minimum energy, it has to overcome a bending energy barrier to reach the minimum energy state. The energy to overcome the barrier can come through appropriate input of external energy, in the absence of which a vesicle can remain in a metastable state.

\subsection{Deformation in a DC field}

Vesicles deform when subjected to a DC electric field as the Maxwell stresses overcome the restoring tension and bending forces. The deformation is predominantly dependent on the conductivity ratio of the inner and outer fluids. To study the dynamics of a vesicle in a DC field, we use an electric field
\begin{equation}
    \mathbf{E}^\infty = E_0 \mathbf{e}_y,\quad t=0:\infty,
\end{equation}
and allow the vesicle to undergo deformation until it reaches a steady shape. Due to the rich dynamics in the deformation of a vesicle at $\sigma_\mathrm{r}=0.1$, our analysis mostly deals with this conductivity ratio.

\subsubsection{Effect of the initial shape}\label{sec:initialshape}
\begin{figure}
    \centering
        \includegraphics[width=0.75\textwidth]{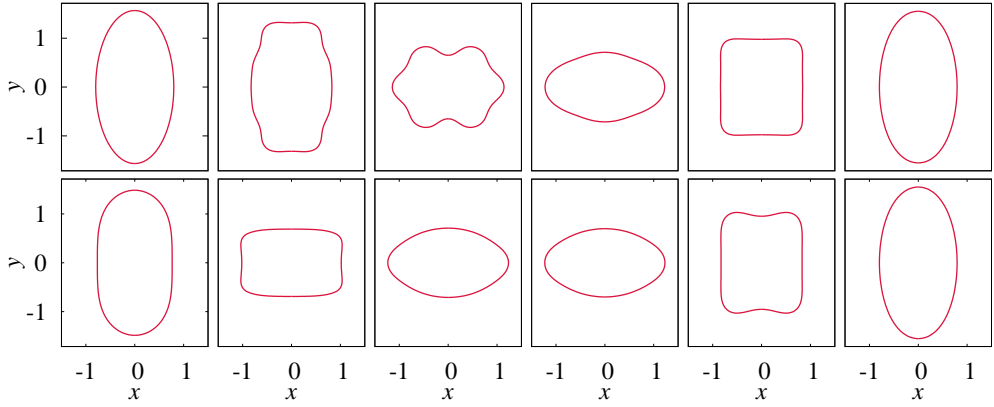}
    \caption{Observed shapes in the deformation of a vesicle with $\nu=0.9$, $\chi=0.02$ in uniform DC electric field at $\beta=4.6$, $\sigma_\mathrm{r}=0.1$, and $\lambda=1.0$, starting with ellipsoid shape (top row) and stress-free shape (bottom row). Shapes correspond to $t=0$, $1.2$, $2.2$, $3.0$, $10.0$, and $t=44.0$.}
    \label{fig:shapesCa4p6}
\end{figure}

An important point to note here is that in the literature,~\cite{ebrahim15,ebrahim15a,Veerapaneni16,mcconnell15sm,mcconnell15} the initial shape of the vesicle has been taken as a prolate ellipsoid such that the prescribed reduced volume is realized. In our simulations, we start with an initial prolate ellipsoid, like earlier studies, but let the vesicle attain an equilibrium shape in the absence of an electric field before switching on the electric field. A vesicle's equilibrium shape is quite different from that of a prolate ellipsoid in the absence of an electric field. Importantly, we observe that starting with such an equilibrium shape of the prolate vesicle (Fig.~\ref{fig:shapesCa4p6}), the  electrohydrodynamic response is quite different from that of a prolate ellipsoid, as considered in the literature.~\citep{ebrahim15a,ebrahim15}  As elaborated in the Appendix, a clear difference is seen in the spatial evolution of TMP, the temporal variation of TMP at the north pole, as well as the normal and tangential electric stresses in the two cases. This indicates that it is crucial to take the correct equilibrium initial shape of the vesicle in the analysis.

The small deformation theory for deflated vesicles does not offer any degree of freedom. In general, the shape of a deformed vesicle can be expressed as $r_s(\theta)=R_o+\Sigma_l s_l P_l(\cos{\theta})$. Here $s_l$ are amplitude coefficients of Legendre modes $P_l(\cos{\theta})$ of order $l$. In the small deformation theory, if one considers only $l=2$ in expansion, corresponding to the Maxwell stress on a sphere at the lowest order, the excess area $\Delta$ is uniquely related to $s_2$, thereby giving only a trivial solution of $s_2=s_{2o}$ and $s_2$, and correspondingly the vesicle shape will not evolve with time. This is really because any change in shape will always invoke higher-order modes.

The constraint of simultaneous volume and area constraint for a deflated vesicle introduces severe nonlinearity as well as necessitates admittance of higher modes of deformation to simultaneously balance normal and tangential stresses at the membrane surface. It can be seen that the shape 2 in the top row of Fig.~\ref{fig:shapesCa4p6} for example can be fitted with $l=2$, $4$ with positive $s_l$'s, while shape 3 in the top row of Fig.~\ref{fig:shapesCa4p6} can be fitted by $l=6$ with a negative $s_l$. The highly deflated vesicle leads to high nonlinearity in curvature stress, whereby higher Legendre modes are excited. These shapes possibly may not be admitted in a droplet where there is the freedom to let the area change. For example, it is known that in droplets, a prolate spheroid ($l=2$, $s_2$ positive) to oblate spheroid ($l=2$, $s_2$ negative) transition happens through an intermediate spherical shape of $s_2=0$. Such a transition necessitates a change in the area, which is permissible in liquid droplets. The same is also possible in quasispherical vesicles in the entropic regime, as discussed in Ref.~\citenum{salipante14}. However, in the case of deflated vesicles, a prolate to oblate transition through intermediate spherical shape is impossible since there is necessarily an area change for the same volume for an intermediate spherical shape. The vesicle, then has only one choice and that is of admitting higher order intermediate shapes to realize such a transformation from prolate to oblate shapes.

\subsubsection{Prolate-oblate-prolate transition in continuous DC fields}
\begin{figure}
    \centering
        \includegraphics[width=0.5\columnwidth]{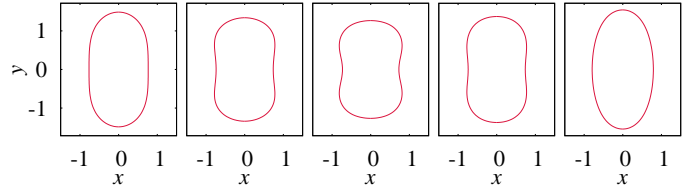}
    \caption{Observed shapes in the deformation of a vesicle with $\nu=0.9$, $\chi=0.01$ in uniform DC electric field at $\beta=0.75$, $\sigma_\mathrm{r}=0.1$, and $\lambda=1.0$.  Shapes correspond to $=0.0$, $3.0$, $7.0$, $12.0$, and $t=44.0$.}
    \label{fig:shapesCa0p75}
\end{figure}

\begin{figure}
    \centering
     \includegraphics[width=0.75\textwidth]{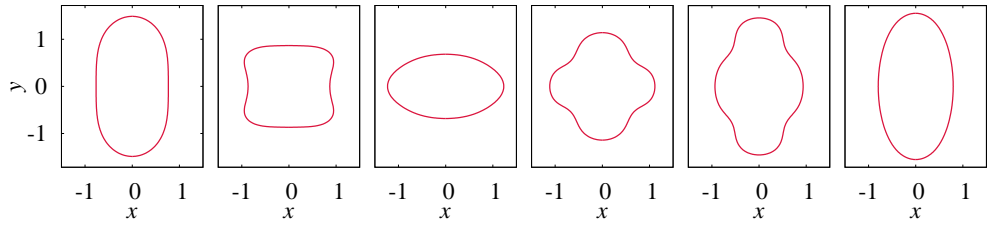}
    \caption{Observed shapes in the deformation of a vesicle in uniform DC field at $\beta=2.0$. Shapes correspond to $t=0$, $2.5$, $4.2$, $10.5$, $13.0$, and $t=44.0$. (Multimedia views)}
    \label{fig:shapesCa2p0}
\end{figure}

An initially prolate vesicle, with a lower conducting inner fluid, in a sufficiently strong uniform DC electric field, undergoes prolate-oblate- transition at short times and an -oblate-prolate transition when the membrane is sufficiently charged, together this is referred to as the prolate-oblate-prolate (POP) transition. For a vesicle with $\nu = 0.9$, $\chi=0.01$, and $\lambda=1.0$, the critical electric field strength for prolate-oblate-prolate transition is $\beta_c=1.01$. For $\beta<\beta_c$, the aspect ratio of a vesicle reduces from its initial aspect ratio for the prolate shape and, finally, recovers a prolate equilibrium shape, without admitting an intermediate oblate shape (Fig.~\ref{fig:shapesCa0p75}). On the other hand, when $\beta>\beta_c$, a clear prolate-oblate-prolate transition is observed [Fig.~\ref{fig:shapesCa2p0} (Multimedia view)].

\begin{figure}
  \centering
    \includegraphics[width=1\textwidth]{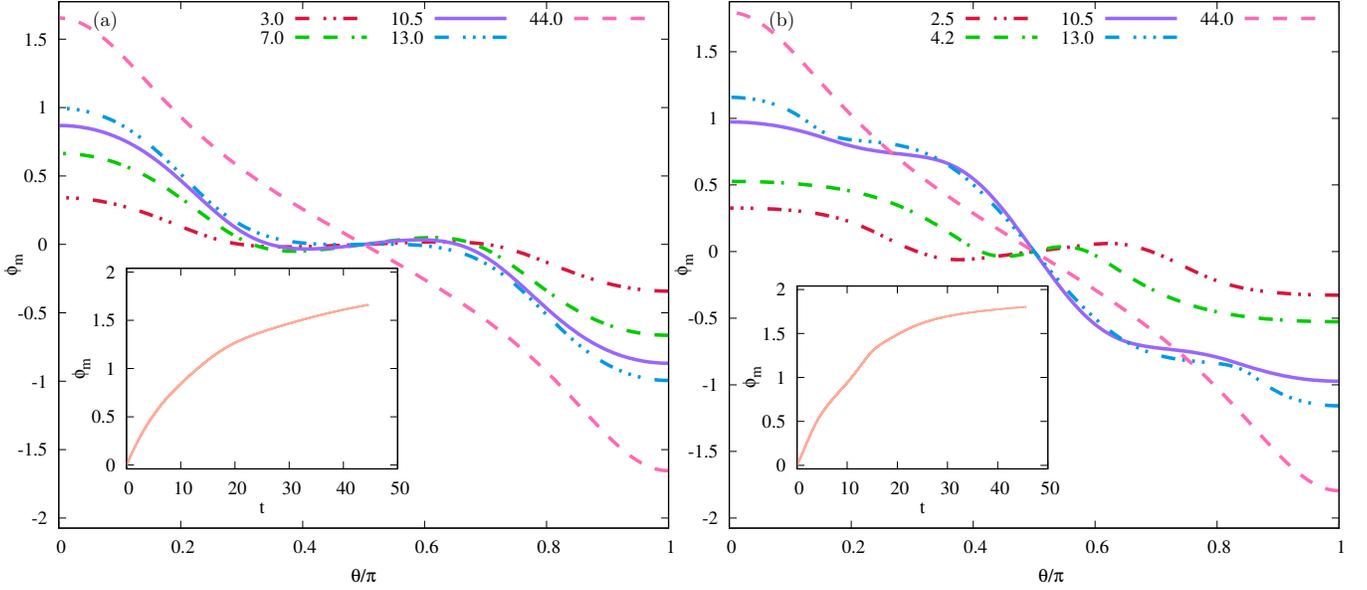}
\caption{The transmembrane potential at the interface at different times and the transmembrane potential at the north pole as a function of time (inset) in the dynamics of a vesicle in a uniform DC field, at (a) $\beta=0.75$ and (b) $\beta=2.0$.}
\label{fig:vmvsthetaca0p75}
\end{figure} 

\begin{figure}
  \centering
    \includegraphics[width=1\textwidth]{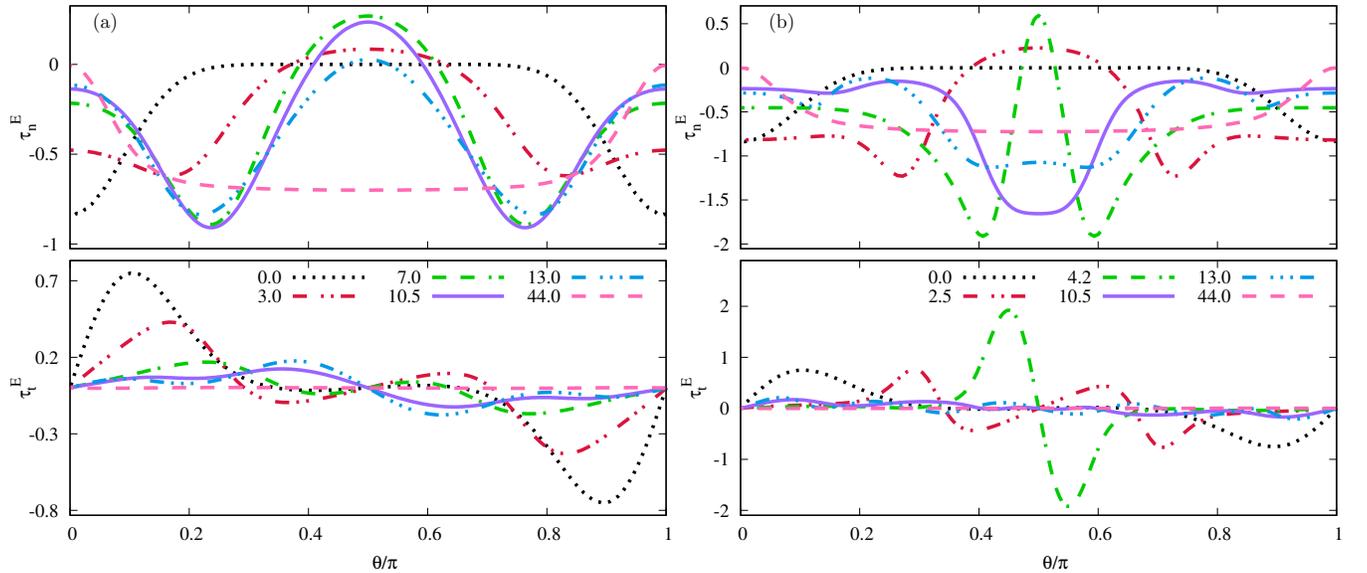}
\caption{Normal and tangential electric stresses at the interface at different times in the dynamics of a vesicle  in a uniform DC field, at (a) $\beta=0.75$ and (b) $\beta = 2.0$.}
\label{fig:tnttca0p75}
\end{figure}

The temporal evolution of the variation of TMP with arclength and the TMP at the north pole, for the two electric field strengths, $\beta=0.75$ [Fig.~\ref{fig:vmvsthetaca0p75}(a)] and $\beta=2$ [Fig.~\ref{fig:vmvsthetaca0p75}(b)], show differences at short times. Although, nearly similar behaviors at long times are observed since the TMP reaches nearly equal values in the fully charged state and similar charging times. However, the normal and tangential electric stress distribution show significant differences [Figs.~\ref{fig:tnttca0p75}(a) and (b)]. While the compressive and tensile normal stresses at the poles and the equator respectively are weak for $\beta=0.75$ ($t=3$, $7$) and cannot overcome the bending stresses, thereby the shape remains prolate; they are quite strong for $\beta=2.0$ ($t=2.5$, $4.2$), favoring intermediate oblate shapes. But importantly, the tangential stresses acting from the poles to the equator, in the case of $\beta=2.0$, are high favoring oblate transition at intermediate times from initial prolate shapes. The normal stresses at a long time then turn highly compressive at the equator ($t=10.5$, $13$) for $\beta=2.0$, which results in transition from intermediate oblate shapes to final prolate shapes. 

\begin{figure}
    \centering
        \includegraphics[width=0.5\textwidth]{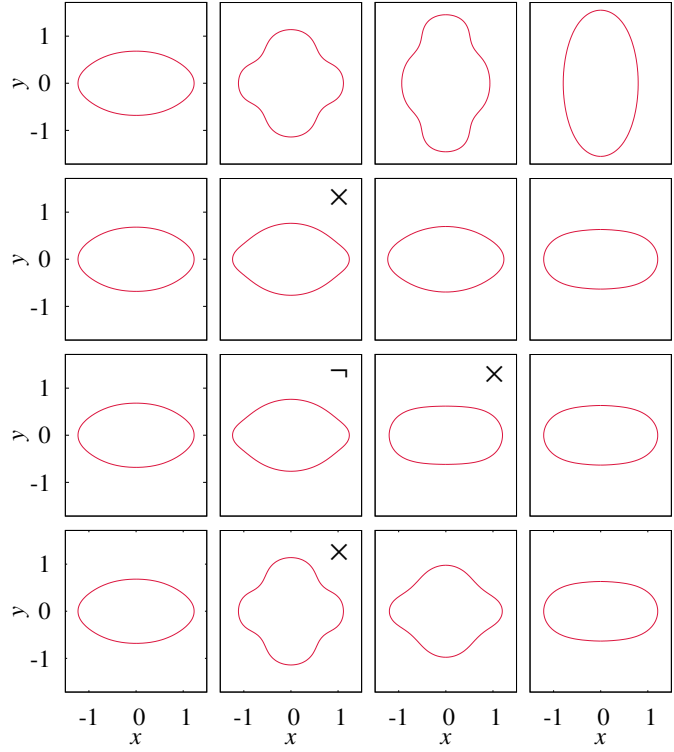}
    \caption{Observed shapes in the deformation of a vesicle in uniform DC field (row $1$, shapes correspond to $t=4.2$, $10.5$, $13.0$, and $t=44.0$), UP($t_\mathrm{m}/2$) (row $2$, shapes correspond to $t=4.2$, $5.25^\times$, $6.0$, and $t=44.0$), BP($t_\mathrm{m}/2 : t_\mathrm{m}/2$) (row $3$, shapes correspond to $t=4.2$, $5.25^\neg$, $10.5^\times$, and $t=44.0$), and UP($t_\mathrm{m}$) (row $4$, shapes correspond to $t=4.2$, $10.5^\times$, $11.0$, and $t=44.0$), at $\beta=2.0$. (Multimedia views)}
    \label{fig:bp}
\end{figure}

\subsection{Deformation in pulsed-DC electric fields}
The dynamics of a vesicle, $\nu=0.9$, $\chi=0.01$, in unipolar and bipolar electric fields of different pulse-widths are studied, setting the dynamics in continuous DC field as the reference. All the results are presented for the system considering $\lambda=1.0$, $\sigma_\mathrm{r}=0.1$, and $\beta=2.0$. We use the notation UP($t_\mathrm{p}$) to indicate a unipolar pulse such that,
\begin{equation}
\begin{split}
    \mathbf{E}^\infty &= E_0 \mathbf{e}_y,\quad t=0:t_\mathrm{p}\\
                          &= 0, \quad t=t_\mathrm{p}:\infty
\end{split}
\end{equation}
while for a bipolar pulse we use BP($t_{p_1}$, $t_{\mathrm{p}_2})$, where
\begin{equation}
\begin{split}
    \mathbf{E}^\infty &= E_0 \mathbf{e}_y,\quad t=0: t_{\mathrm{p}_1}\\
                          &= -E_0\mathbf{e}_y, \quad t=t_{\mathrm{p}_1} : t_{\mathrm{p}_2}\\
                          &= 0, \quad t=t_{\mathrm{p}_2} : \infty
\end{split}
\end{equation}

\begin{figure}
    \centering
\includegraphics[width=1\textwidth]{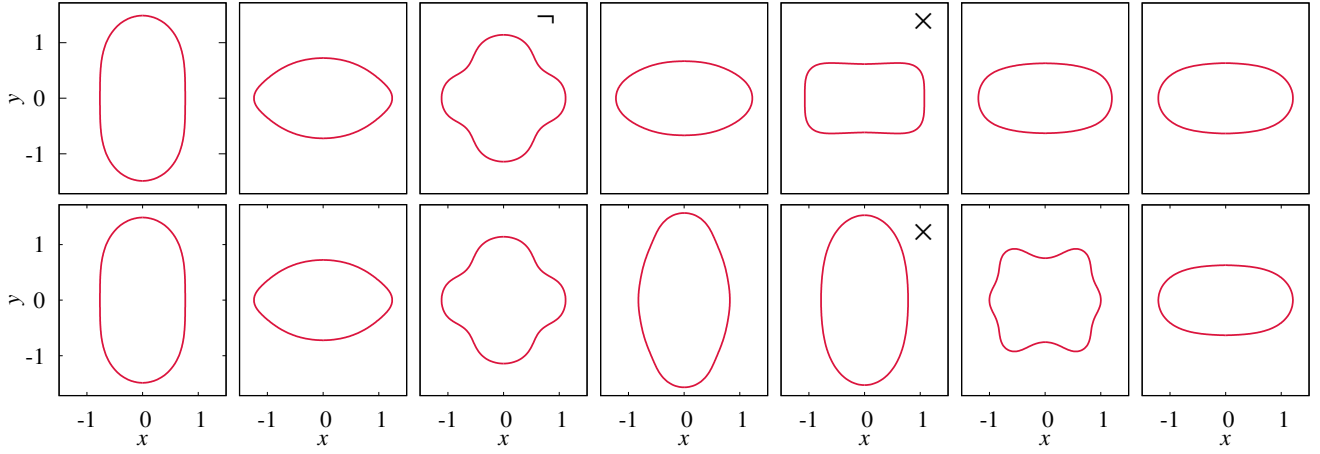}
    \caption{Observed shapes in the deformation of a vesicle in (top row) BP($t_\mathrm{m}$, $t_\mathrm{m}$) and (bottom row) UP($2t_\mathrm{m}$) at $\beta=2.0$. Shapes correspond to $t=0$, $4.5$, $10.5^\neg$, $11.0$, $21.0^\times$, $22.0$, and $t=44.0$ for BP($t_\mathrm{m}$, $t_\mathrm{m}$) and $t=0$, $4.5$, $10.5$, $15.0$, $21.0^\times$, $25.0$, and $t=44.0$ for UP($2t_\mathrm{m}$).}
    \label{fig:bp2}
\end{figure}

\begin{figure}
  \centering
    \includegraphics[width=0.5\textwidth]{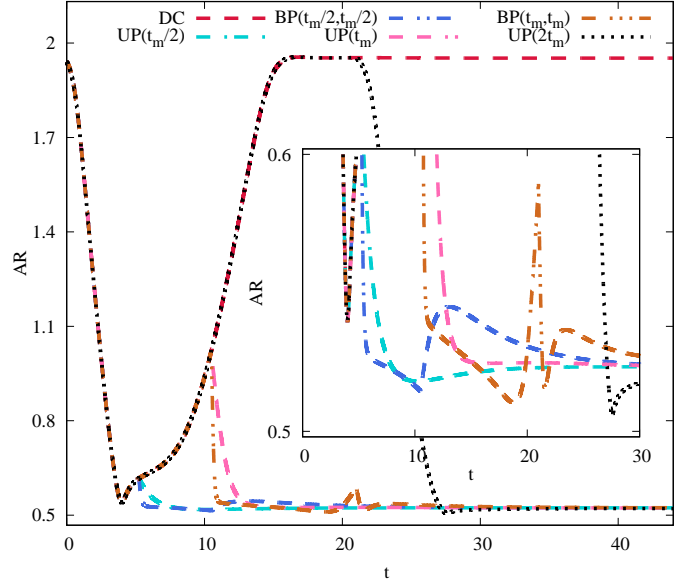}
\caption{In the dynamics of a vesicle, aspect ratios as the function of time in different types of fields at $\beta=2.0$ and $\sigma_\mathrm{r}=0.1$. Inset shows the zoomed-in plot of the same.}
\label{fig:arvst}
\end{figure} 

A few shapes are marked with symbols in the dynamics of a vesicle in unipolar and bipolar pulsed-DC fields. The symbol ``$\neg$'' indicate the time at which the direction of the bipolar pulsed-DC field changes, symbol ``$\times$'' indicates the time at which the pulse ends, \ie,  the field is turned off, and symbol ``$\dagger$'' indicate the time at which the field strength is reduced.

Figure~\ref{fig:bp} (Multimedia view) shows the evolution of vesicle shape in (i) continuous DC (redrawn to compare), (ii) UP($t_\mathrm{m}/2$), (iii) BP($t_\mathrm{m}/2$, $t_\mathrm{m}/2$), (iv) UP($t_\mathrm{m}$), and Fig.~\ref{fig:bp2} for (v) BP($t_\mathrm{m}$, $t_\mathrm{m}$) and (vi) UP($2 t_\mathrm{m}$). These figures indicate the sensitive dependence of intermediate shapes on the type of pulse and the pulse width. For these cases, the aspect ratios as the functions of time are plotted in Fig.~\ref{fig:arvst}. The shapes before $t=4.2$, shown in Fig.~\ref{fig:shapesCa2p0}, are excluded in Fig.~\ref{fig:bp}.  
Figures.~\ref{fig:bp} and~\ref{fig:bp2} show that for $\beta=2.0$, except continuous DC where the final shape is a prolate spheroid, all the other pulsed-DC fields lead to  final oblate spheroid shapes. This is even more puzzling for the case of UP($2 t_\mathrm{m}$), where, while the vesicle has attained a prolate shape at the end of the pulse, switching off the pulse leads to it being \emph{pushed} into an oblate shape. Thus, for $\beta=2$, $\sigma_\mathrm{r}=0.1$, irrespective of the intermediate shape, at the end of the pulse, the final shape seems to be an oblate.

\begin{figure}
  \centering
   \includegraphics[width=1\textwidth]{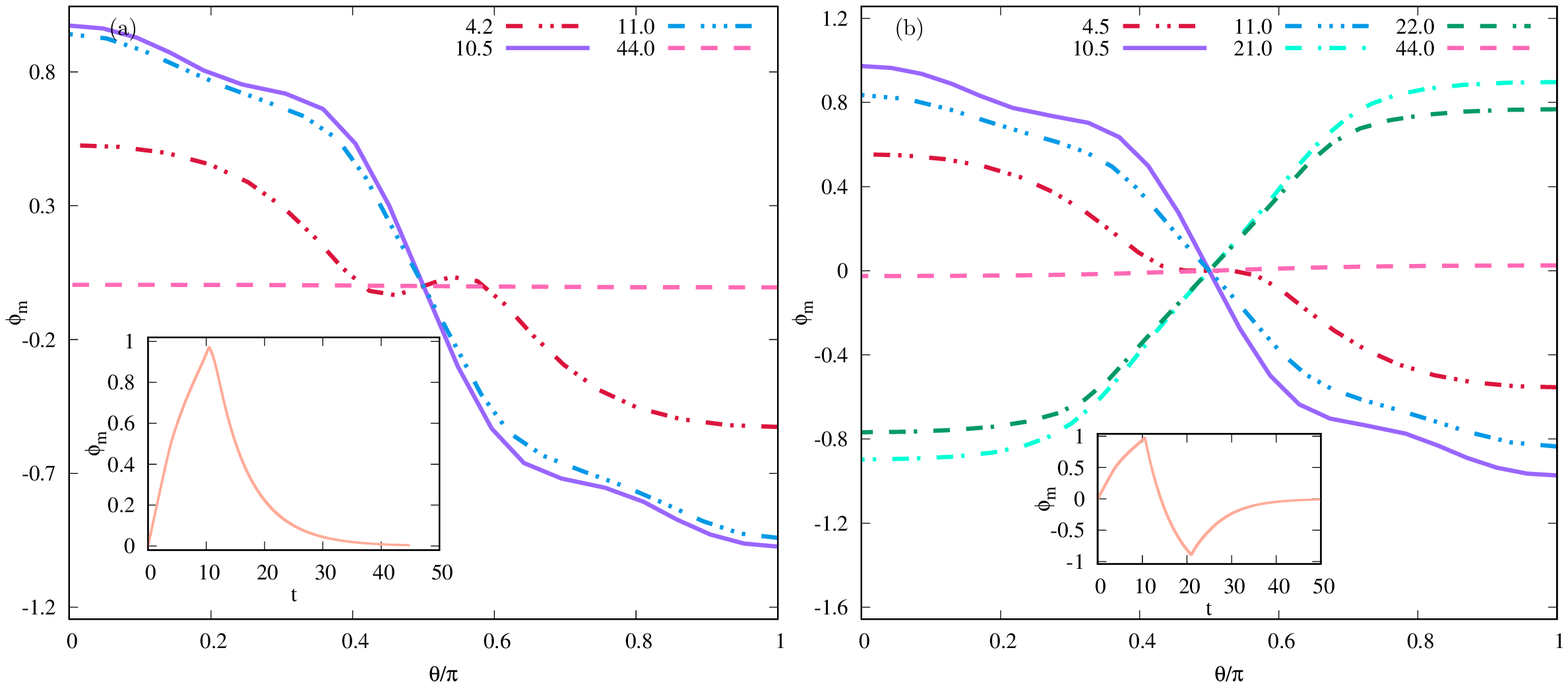}
\caption{The transmembrane potential at the interface at different times and the transmembrane potential at the north pole as a function of time (inset) in the dynamics of a vesicle at $\beta=2.0$ in (a) UP($t_\mathrm{m}$) (b) BP($t_\mathrm{m}$, $t_\mathrm{m}$).}
\label{fig:vmvsthetaca2p0bt}
\end{figure}

It is useful to compare UP($t_\mathrm{p}$) and BP($t_\mathrm{p},t_\mathrm{p}$) since often a bipolar pulse is introduced to hasten the discharge of a vesicle by inverting the TMP, as seen in the TMP plots (\eg, Figs.~\ref{fig:vmvsthetaca2p0bt}(a) and~\ref{fig:vmvsthetaca2p0bt}(b) for $t_\mathrm{p}=t_\mathrm{m}$, and Figs.~\ref{fig:vmvsthetaca2p0bth}(a) and~\ref{fig:vmvsthetaca2p0bth}(b) in the Appendix for $t_\mathrm{p}=t_\mathrm{m}/2$). Consider the evolution of a vesicle in UP($t_\mathrm{m}$) (shown in the last row of Fig.~\ref{fig:bp}) and BP($t_\mathrm{m}$, $t_\mathrm{m}$) (shown in the first row of Fig.~\ref{fig:bp2}).  A comparison of the TMP evolution for UP($t_\mathrm{m}$) and BP($t_\mathrm{m}$, $t_\mathrm{m}$) shows that the TMP indeed reverses in the case of BP, and faster decay of TMP is observed. The comparison of shape analysis for UP($t_\mathrm{m}$) and BP($t_\mathrm{m}$, $t_\mathrm{m}$) also shows significant differences. The UP($t_\mathrm{m}$) pulse shows an equatorial torus (at $t=t_m=10.5$) at the end of the pulse, before evolving back into an oblate spheroid (Fig.~\ref{fig:vmvsthetaca2p0bt}). On the other hand, in the case of the BP pulse, the equatorial torus evolves into an oblate spheroid, but through an intermediate shape that is an oblate cylinder ($t=21$) (Fig.~\ref{fig:bp2}). Interestingly, the final shape is the same oblate shape as for the UP ($t_m$) when the pulse is switched off. 

\begin{figure}
  \centering
    \includegraphics[width=1\textwidth]{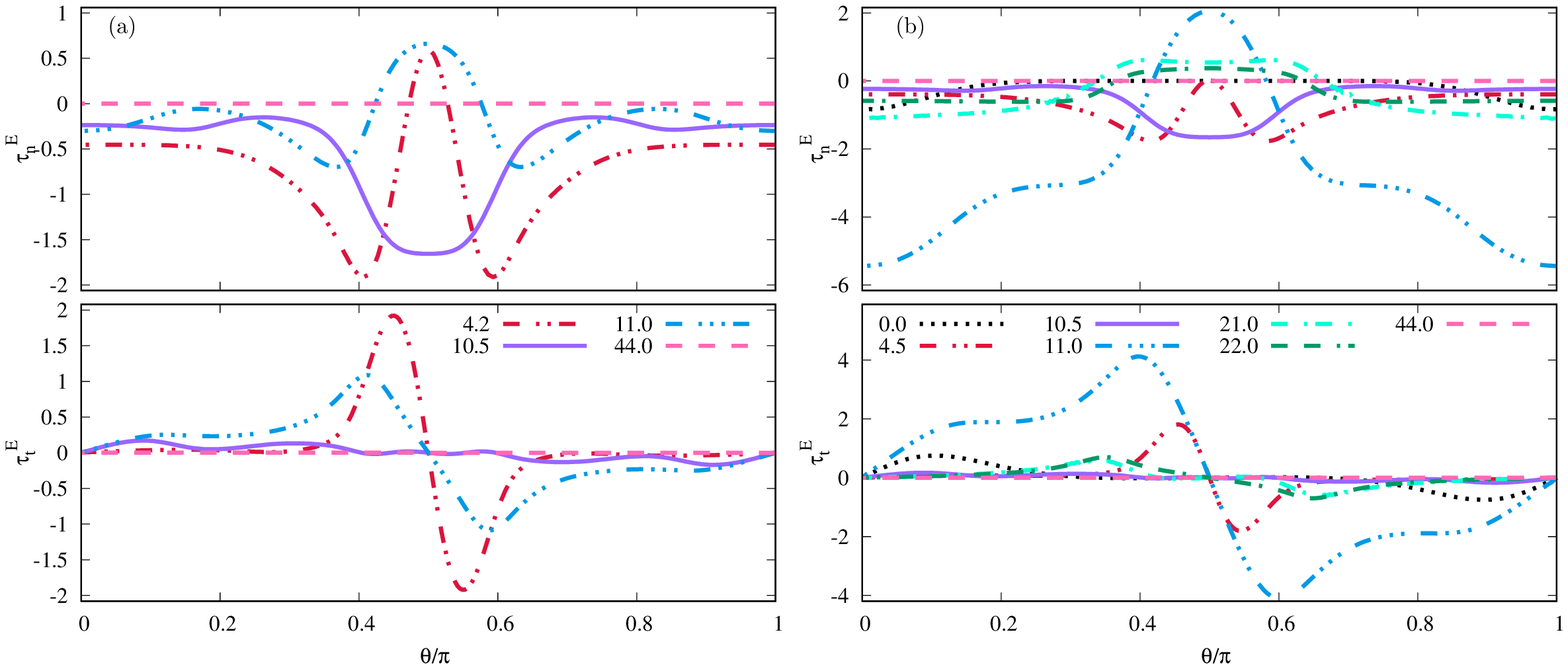}
\caption{Normal (top) and tangential (bottom) electric stresses at the interface at different times in the dynamics of a vesicle at $\beta=2.0$ in (a) UP($t_\mathrm{m}$) (b) BP($t_\mathrm{m}$, $t_\mathrm{m}$).}
\label{fig:tnttca2p0bpts}
\end{figure}

It is also useful to compare the dynamics in the UP($t_\mathrm{p}$) with that of in the BP($t_\mathrm{p}/2$, $t_\mathrm{p}/2$) as the total pulse times and energy inputs are the same. A bipolar pulse BP($t_\mathrm{m}/2$, $t_\mathrm{m}/2$) is found to hasten the evolution to oblate shape as compared to that of a unipolar pulse of the same total pulse time UP($t_\mathrm{m}$) (Fig.~\ref{fig:bp}) or admit an additional oblate shape (at $t=11.0$) for BP($t_m,t_m$) wherein UP($t_m$) does not (Fig.~\ref{fig:bp2}). The reversal of polarity in bipolar pulse for a given total pulse time leads to a reversal of stresses (Fig.~\ref{fig:tnttca2p0bpts}) and corresponding shape. The intermediate -oblate-prolate- transition is completely prevented in the case of the bipolar pulse, such that prolate shapes are excluded at long times. 

\begin{figure}
  \centering
   \includegraphics[width=0.5\textwidth]{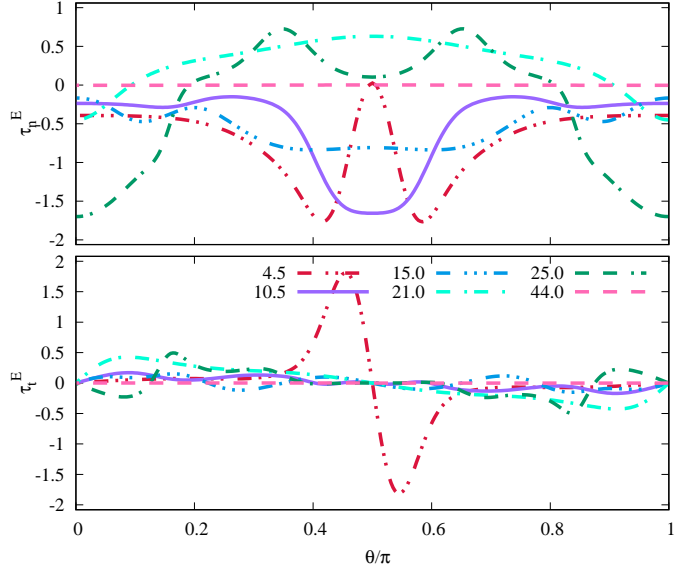}
\caption{Normal (top) and tangential (bottom) electric stresses at the interface at different times in the dynamics of a vesicle at $\beta=2.0$ in UP($2t_\mathrm{m}$).}
\label{fig:tnttca2p0bp2t}
\end{figure} 

The mechanism of prolate-oblate-prolate-oblate transition in vesicles subjected to strong DC-pulses, when the outer medium is more conducting, can be understood from the electric stress distribution, shown in Figs.~\ref{fig:tnttca2p0bpts} and \ref{fig:tnttca2p0bp2t}. The normal stress distribution on the initially prolate vesicle is compressive at the poles, which facilitates prolate-oblate- transition. This continues, and the strong compressive stresses at $45^{o}$ (at $t=4.5$) promote oblate cylindrical shapes, which then relax due to fairly homogenous but moderate compressive stresses throughout the equatorial region ($t=15.0$), which promote prolate shapes (Fig.~\ref{fig:tnttca2p0bp2t}). When the field is switched off at $t=2 t_\mathrm{m}=21$, a sudden redistribution of stresses, on account of discharge of the capacitor membrane, occurs, such that the stresses are now tensile in the equatorial region. The tensile normal equatorial stresses, supported by tangential stresses, which act from poles to the equator, lead to a prolate-oblate-prolate-oblate transition. When the membrane is completely discharged, the Maxwell stresses decrease to zero, and the vesicle is trapped in a metastable oblate shape.

The generation of tensile equatorial stresses, when the pulse is switched off, can be understood as follows. The equatorial normal stress is given by $\frac{1}{2}\left[ (E_{n,\mathrm{e}}^2-E_{n,\mathrm{i}}^2)-(E_{t,\mathrm{e}}^2-E_{t,\mathrm{i}}^2) \right]$. When the membrane is fully charged, the normal electric field is zero both at the inner and outer sides, \ie, $E_{n,\mathrm{i}}=E_{n,\mathrm{e}}=0$. Similarly, $E_{t,\mathrm{i}}$ is also zero, whereby the normal stresses are compressive at the equator due to non-zero $E_{t,e}$, which promotes prolate spheroidal shape. However, when the capacitor discharges, on switching off the pulse, while $E_{n,\mathrm{e}}=E_{n,\mathrm{i}}=0$, the tangential electric field at the inner side $E_{t,\mathrm{i}}$ is not zero. in fact, owing to the higher conductivity of the outer fluid, $E_{t,\mathrm{i}}>E_{t,\mathrm{e}}$ since the charges on the outer side of the membrane dissipate faster. This leads to tensile normal electric stresses at the equator during the discharge of the capacitor, resulting in oblate shapes.

It is then interesting to compare the stress distribution in the cases UP($t_\mathrm{p}$) and BP($t_\mathrm{p}/2$, $t_\mathrm{p}/2$). For $t_\mathrm{p}=2 t_\mathrm{m}$, we find that the stress distribution is similar up to $t=t_\mathrm{m}$  (Figs.~\ref{fig:tnttca2p0bpts} and \ref{fig:tnttca2p0bp2t}).  Thereafter, while in the UP case, the compressive stresses at the equator result in an -oblate-prolate- transition ($t=15$), in the case of BP, a dramatic stress reversal is observed at $t=11$,  such that the equatorial stresses are tensile. A bipolar pulse sets in an immediate discharging of the vesicle such that the membrane charge now equilibrates to the reversed polarity of the bipolar pulse. This results in a mechanism similar to the discharging of a charged capacitor discussed earlier, hastening the onset of tensile equatorial stresses. The early onset of the pole to equatorial tangential stresses can be seen in BP($t_\mathrm{m}$, $t_\mathrm{m}$) at $t=11$, unlike that in UP($t_\mathrm{m}$).

While comparing shape changes for UP($t_\mathrm{p}$) and BP($t_\mathrm{p}$, $t_\mathrm{p}$), \eg, $t_\mathrm{p}=t_\mathrm{m}$, it is seen in Figs.~\ref{fig:tnttca2p0bpts}(a) and~\ref{fig:tnttca2p0bpts}(b) that the stress distribution favors an -oblate-prolate- transition at the end of $t=t_\mathrm{m}=10.5$. In the case of a unipolar pulse ($t_\mathrm{p}=t_\mathrm{m}$), the discharging of the capacitor ($t=11.0$) generates equatorial tensile stresses pushing the vesicle to an oblate shape. The compressive stresses at $\theta=45^{o}$ result in a typical equatorial ring-shaped vesicle. In the bipolar case, though, the strong equatorial tensile stress and strong compressive stress at the poles result in an oblate shaped vesicle at the end of the bipolar pulse.

\subsubsection{Deformation in two-step pulsed-DC electric field}

Although numerical calculations predicted the prolate-oblate-prolate transitions in time,~\cite{mcconnell13} only the prolate-oblate transition was seen in experiments.~\citep{karin06} The -oblate-prolate transition at long times, as predicted by simulations,~\citep{mcconnell15sm} was never observed in unipolar pulses.~\citep{boukany17} \citet{salipante14} argued that this is essentially because a very strong DC field is required for admitting the prolate-oblate- transition. This field typically leads to simultaneous electroporation of the membrane. An electroporated membrane, which has finite membrane conductance, is like a liquid droplet, wherein when the outer conductivity is higher, an oblate shape is observed. Such a membrane never gets fully charged, though the -oblate-prolate transition is not seen in experiments. To circumvent the electroporation, they theoretically designed and experimentally implemented a two-step pulse algorithm, wherein a short-strong-pulse causes the prolate (or sphere)-oblate- transition. The shortness of this pulse prevents the membrane poration. A weak DC-pulse in the series leads to charging of the vesicle, and thereby inducing the -oblate-prolate transition at long time. This weak second pulse ensures a prolate shape at the end of the second pulse (seen in both experiments and theory).


The Schwan equation, $\phi_\mathrm{m}=\frac{3}{2} a E_o  \left[1-\exp(-t_\mathrm{sp}/t_\mathrm{m})\right]$, correlates the potential induced by an external field across the membrane of a spherical cell.~\citep{schwan1983,schwan92} In the two-step pulse analysis, to circumvent the membrane electroporation, the pulse width of the strong short pulse $t_\mathrm{sp}$ has to be such that the $\phi_\mathrm{m}$ does not exceed $1$~V. This meant that the strong pulse had to be really short. Simultaneously, it had to be ensured that $t_\mathrm{sp}\sim t_\mathrm{h}$, such that the hydrodynamic response is quick enough. The weak pulse was then applied for a time $t_\mathrm{wp}>t_\mathrm{m}$. Their study indicated that the long-weak-pulse had to have $\beta<\beta_c$. Therefore, they reduced the duration of the strong pulse by having a short-strong-pulse, followed by a long-weak-pulse, with the pulse sign kept the same.

\begin{figure}
\begin{center}
\includegraphics[width=1\textwidth]{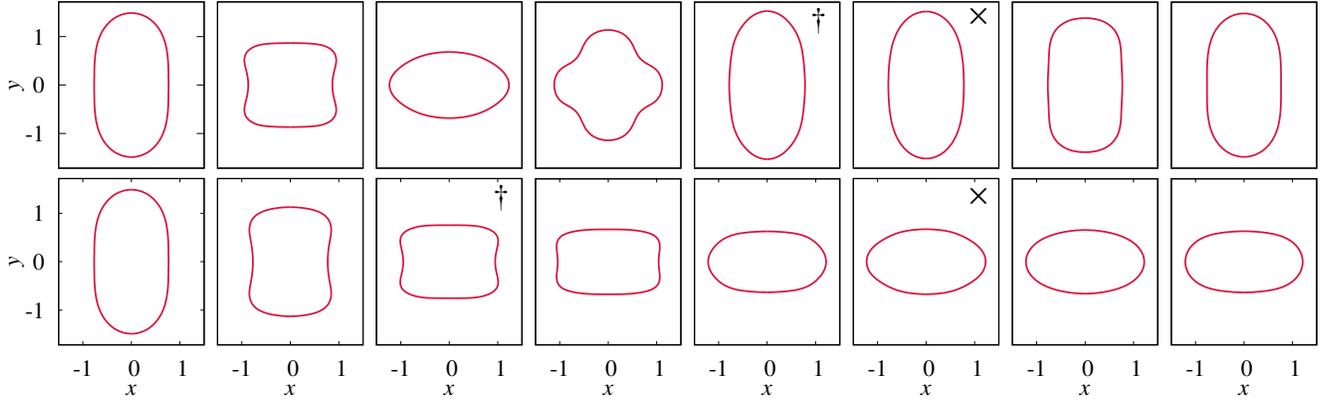}
\caption{Deformation of a vesicle in (top row) UP($2t_\mathrm{m}$) and (bottom row) UP($t_\mathrm{m}$). In UP($2t_\mathrm{m}$), the field strength is reduced at $t=17$ from $\beta_1=2.0$ to $\beta_2=0.7$, and the shapes are shown for $t=0$, $2.5$, $4.2$, $10.5$, $17^\dag$, $21^\times$, $32$, $\infty$. In UP($t_\mathrm{m}$), the field strength is reduced at $t=3$ from $\beta_1=2.0$ to $\beta_2=0.7$, and the shapes are shown for $t=0$, $1.5$, $3.0^\dag$, $4.2$, $7$, $10.5^\times$, $12$, $\infty$.}
\label{fig:fieldshift}
\end{center}
\end{figure}

The vesicle dynamics in a two-step pulse is simulated in our work in two ways. In one case, at first, a long-strong-pulse is provided, which causes prolate-oblate-prolate transition in time, resulting in a prolate deformation at the end of this strong pulse. In the series, a weak second pulse is applied before turning the pulse off (top row in Fig.~\ref{fig:fieldshift}). In this case, the normal stresses are compressive at the equator since $t>t_\mathrm{m}$, at the end of the long-strong-pulse. Further, when the electric field strength is reduced, these stresses reduce significantly, such that when finally switched off, at $t=2 t_\mathrm{m}$, the tensile normal stresses at the equator are not enough to pull it back into an oblate state. Therefore, the final equilibrium shape of the vesicle is prolate.

In another simulation, we first apply a short-strong-pulse, which causes a prolate-oblate transition in time. Whereafter a second weak-pulse is applied, the resultant weak compressive normal stresses at the equator are unable to cause an oblate-prolate transition. The vesicle, therefore, remains in the oblate shape (bottom row in Fig.~\ref{fig:fieldshift}). When the field is switched off, the weak tensile equatorial stresses further assist the oblate shape, and the vesicle remains in the oblate state.

Our results can be contrasted with the results reported in Ref.~\citenum{salipante14} for the dynamics of a vesicle in two-step pulsed electric field, which argues that a short-strong-pulse followed by a long-weak-pulse can induce -oblate-prolate transition in time. This is really due to the assumption of a quasispherical vesicle (which in the model is captured by thermal fluctuations and available excess area) wherein the bending energy barrier to -oblate-prolate transition is not felt. This is also apparent from their calculations and experiments since the vesicles return to deformation $D\approx 1$, which indicates a spherical (quasispherical) shape. In this limit, our results predict that an -oblate-prolate-oblate transition can be prevented by a long-strong-pulse followed by a short-weak-pulse.

\subsubsection{Deformation of a vesicle with the more conducting inner fluid}
\begin{figure}
    \centering
\includegraphics[width=0.5\textwidth]{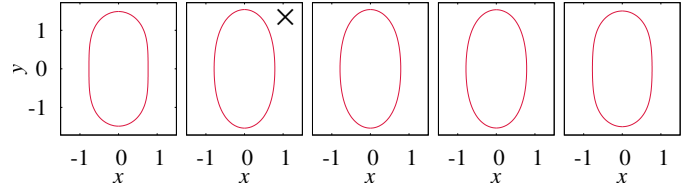}
    \caption{Observed shapes in the deformation of a vesicle in UP($t_\mathrm{m}$) at $\beta=2.0$ and $\sigma_\mathrm{r} = 10$. Shapes correspond to $t=0$, $0.6^\times$, $0.65$, $1.0$, and $18.0$.}
    \label{fig:shapesCa2p0sigr10}
\end{figure}

The dynamics of a vesicle in UP($t_\mathrm{m}$) at $\sigma_\mathrm{r}=10$ is shown in Fig.~\ref{fig:shapesCa2p0sigr10}. In this case, the stresses for $t<t_\mathrm{m}$ are tensile at the poles favoring prolate shapes. For $t>t_\mathrm{m}$, the stresses are compressive at the equator, again favoring prolate shape. Thus the vesicle remains prolate spheroid when the field is \emph{on}. When the pulse is switched off, the normal electric stresses indeed become tensile at the equator. However, the electric stresses at the poles are also tensile and of much higher magnitude, resulting in an equilibrium prolate configuration after the discharging of the membrane.

\section{Conclusions} \label{Concl}
A significant effect of a DC-pulse, in general and bipolar pulse, in particular, is numerically predicted for a vesicle subjected to strong DC fields, especially for $\sigma_\mathrm{r}<1$. Switching off the field leads to tensile equatorial stresses, which push the vesicle into a metastable oblate spheroidal shape. The bipolar pulses hasten and assist this process again due to strong tensile equatorial stresses, generated due to polarity reversal.
Additionally, this analysis suggests that the recent use of bipolar pulses for TMP regulation also leads to the significant corresponding change in the electrodeformation of vesicles and thereby the spatial (surface) distribution of TMP.

The typical values of dimensional parameters to study vesicle dynamics in typical experiments as inspired by the present work would be $a\approx 10\ \mu$m,  $\mu_\mathrm{e} \approx 10^{-3}$~Pa s, $\sigma_\mathrm{e}\approx 1$--$10\times 10^{-5}$~S m$^{-1}$, $\epsilon_\mathrm{e} \approx 80$, $\kappa_\mathrm{b} \approx 10^{-19}$~J, and $C_\mathrm{m} \approx 10^{-2}$ F m$^{-2}$. With this set of parameters, the field strength seems reasonable \{$E_0 = \left[\beta \sigma_\mathrm{e}\mu_\mathrm{e}/(a C_\mathrm{m}\epsilon_\mathrm{e}\epsilon_0)\right]^{1/2} \approx 0.1$--$1$ kV cm$^{-1}$\}, whereas the considered value of dimensionless bending stiffness $\chi= 0.01$ seems to be a little high. Consideration of a lower value $\chi < 0.01$ causes numerical failure; this is one drawback in our numerical method. A high value of bending rigidity is, however,  commonly used in the earlier computational analyses.~\citep{mcconnell15sm,veerapaneni19pair} Our numerical studies indicate that in a strong electric field, the consideration of a slightly higher value of $\chi$ does not significantly change the dynamics of the deformation of a vesicle. Therefore, the analysis presented in this article can be demonstrated in an experiment.  

This study, although motivated by electroporation, does not consider pore formation in the membrane and can be considered, together with similar such studies in literature, as steps toward eventually describing electroporation with electrodeformation. Our work presents the variation of TMP along the arclength of the membrane as a function of time, as well as with different pulse parameters (Figs.~\ref{fig:vmvsthetaca0p75}, \ref{fig:tnttca0p75} and other figures showing TMP vs arclength and time). The Schwan model predicts a $\cos{\theta}$ variation of TMP such that it is maximum at the poles and disappears at the equator. Our study indicates two important things; first, it demonstrates that the TMP variation with arc length can change significantly as deformation proceeds after the electric field is switched off. Thus, a considerably larger surface of the cell membrane can get porated. Second, it indicates that the temporal control of TMP, and thereby the spatial control, can be attained by the use of bipolar pulses. The exact problem of coupling electrodeformation with electroporation is part of our future study in this direction.
 
\begin{acknowledgments}
We acknowledge the financial support from Labex MEC (grant no. ANR-11-LABX-0092) and ANR (grant no. ANR-18-CE06-0008-03). Centre de Calcul Intensif d'Aix-Marseille is acknowledged for granting access to its high-performance computing resources.
 
\end{acknowledgments}

 \appendix*\label{appendix}
\section{Additional figures and the accuracy of the code}

 \begin{figure}
  \centering
    \includegraphics[width=1\textwidth]{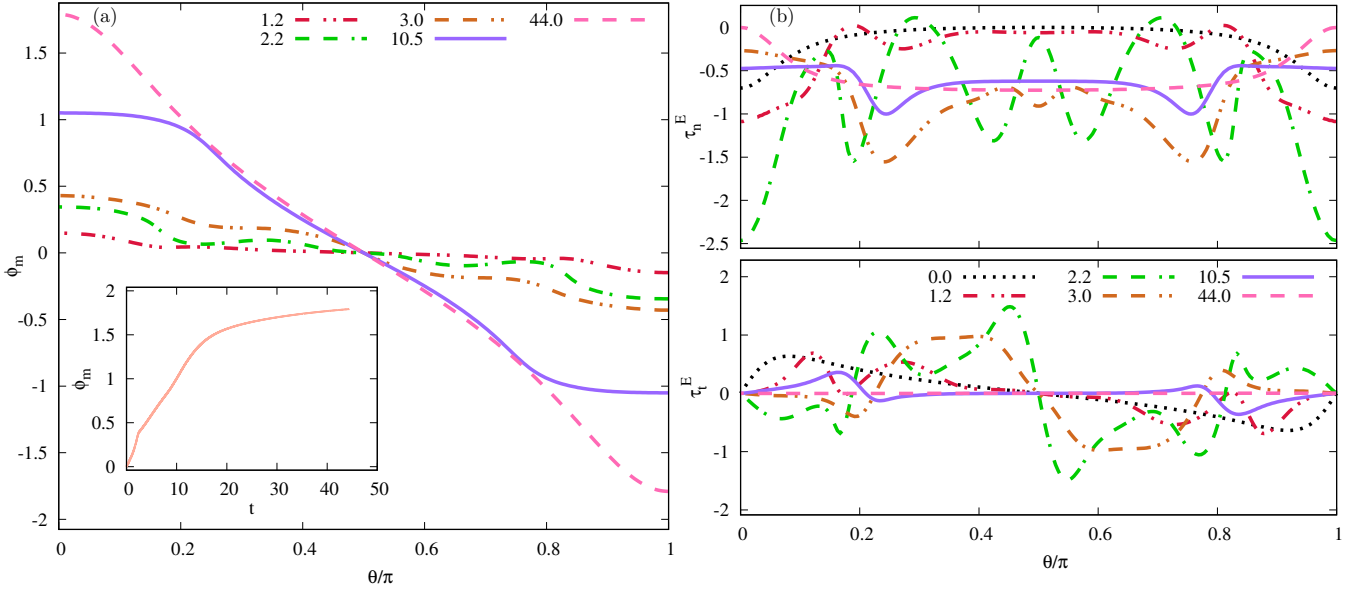}
\caption{in the dynamics of a vesicle with initial ellipsoid shape at $\beta=4.6$ in a DC field, (a) the transmembrane potential at the interface at different times and the transmembrane potential at the north pole as a function of time (inset), (b) normal and tangential electric stresses at the interface at different times. }
\label{fig:tnttca4p6ab}
\end{figure}

\subsection{Additional data and figures}\label{appendixA}

\begin{figure}
  \centering
   \includegraphics[width=1\textwidth]{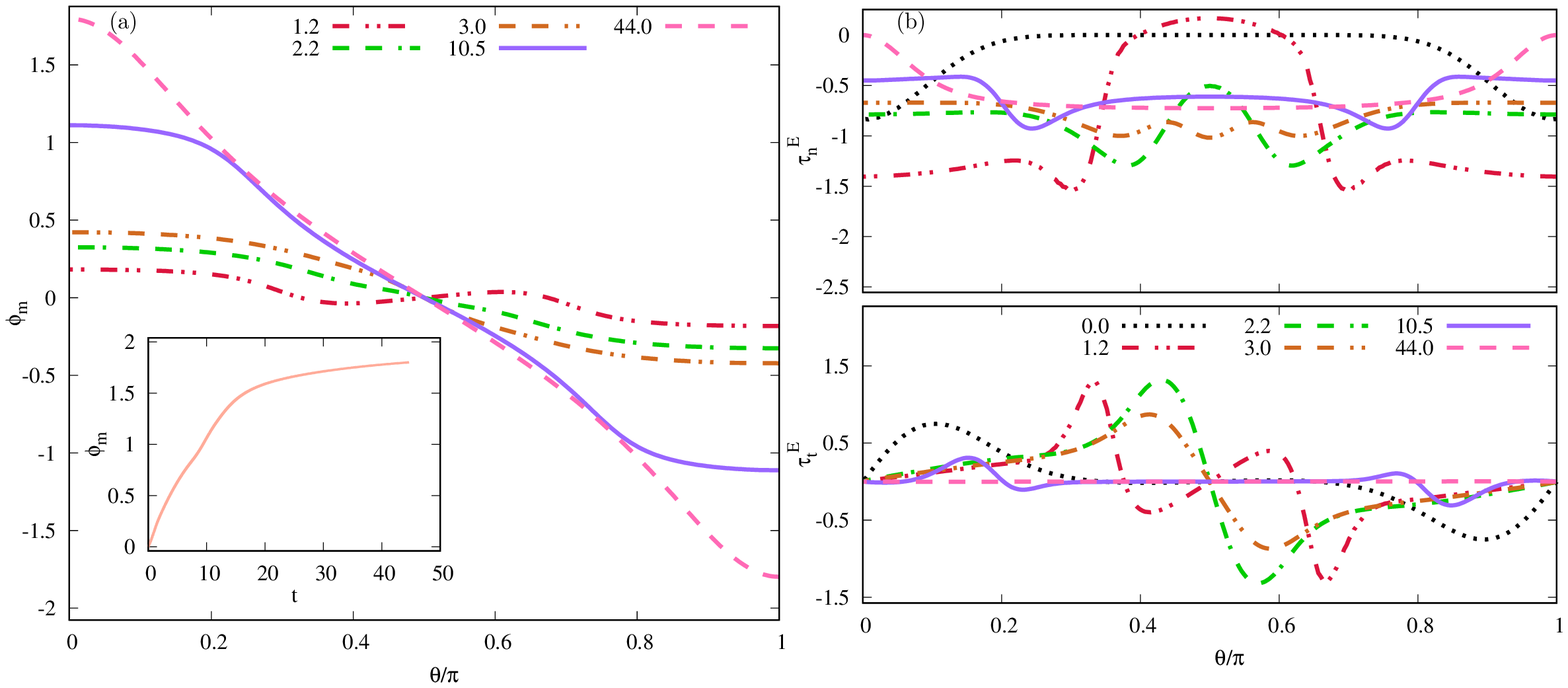}
\caption{In the dynamics of a vesicle with initial stress-free shape at $\beta=4.6$ in a DC field, (a) the transmembrane potential at the interface at different times and the transmembrane potential at the north pole as a function of time (inset), (b) normal and tangential electric stresses at the interface at different times.}
\label{fig:tnttca4p6eq}
\end{figure}

\begin{figure}
  \centering
   \includegraphics[width=1\textwidth]{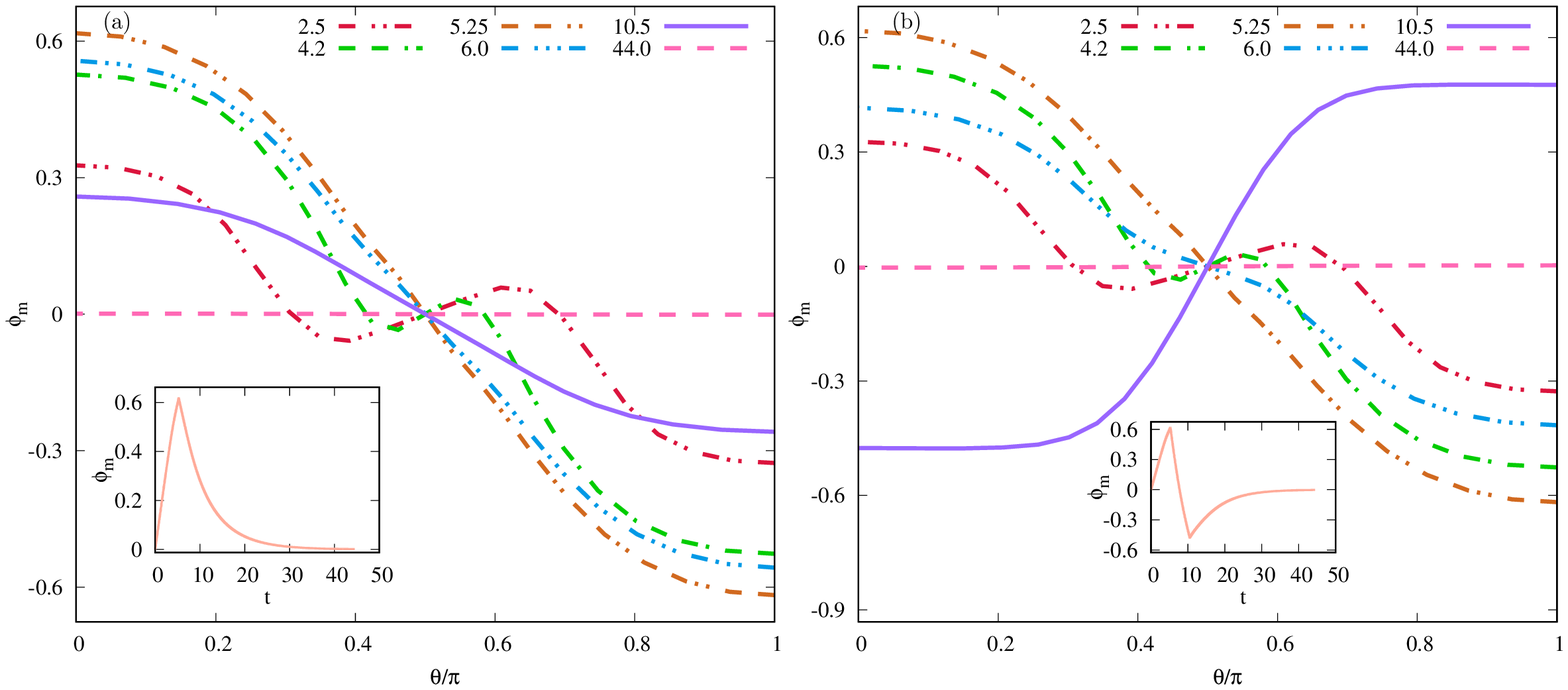}
\caption{The transmembrane potential at the interface at different times and the transmembrane potential at the north pole as a function of time (inset) in the dynamics of a vesicle at $\beta=2.0$ in (a) UP($t_\mathrm{m}/2$) (b) BP($t_\mathrm{m}/2$, $t_\mathrm{m}/2$).}
\label{fig:vmvsthetaca2p0bth}
\end{figure}

The dynamics of a vesicle in a DC electric field depends upon the consideration of the initial shape, as discussed in Sec.~\ref{sec:initialshape}. Figure~\ref{fig:tnttca4p6ab}(a) shows the variation of transmembrane potential at the interface, and Fig.~\ref{fig:tnttca4p6ab}(b) shows the electric stresses at the interface at different times for the dynamics of a vesicle with an initial prolate ellipsoid shape. For this case, the evolution of the transmembrane potential at the north pole with time is shown in the inset of Fig.~\ref{fig:tnttca4p6ab}(a). Similarly, in the dynamics of a vesicle with an initial stress-free shape, the transmembrane potential at the interface, and the electric stresses at different times are shown in Figs.~\ref{fig:tnttca4p6eq}(a) and~\ref{fig:tnttca4p6eq}(b), respectively. For this case as well, the evolution of the transmembrane potential at the north pole with time is shown in the inset of Fig.~\ref{fig:tnttca4p6eq}(a).

In the electrodeformation of vesicles in UP($t_\mathrm{m}/2$) (shown in the second row of Fig.~\ref{fig:bp}) and BP($t_\mathrm{m}/2$, $t_\mathrm{m}/2$) (shown in the third row of Fig.~\ref{fig:bp}), a comparison of the TMP evolution in Figs.~\ref{fig:vmvsthetaca2p0bth}(a) and~\ref{fig:vmvsthetaca2p0bth}(b) shows that the TMP clearly falls faster for the BP pulse, changing sign and thereby decaying to zero. It can also be observed that the TMP reverses in the case of BP, and it decays faster than the case of BP($t_\mathrm{m}/2$, $t_\mathrm{m}/2$). 

\subsection{Optimization of the boundary integral code}\label{appendixB}
For solving the dynamics of a vesicle in an electric field, the maximum time-step size is $\Delta t = 10^{-5}$, which allows the computation with highly accurate results. With this time-step size, the performance of the code depends upon the number of elements. Further lowering the time-step size leads to the very slow performance of the code, which makes the computation impossible to carry out. Therefore, $\Delta t = 10^{-5}$ is considered, and the performance of the code is optimized by selecting appropriate number of elements, which is given in table~\ref{tab:table1}. 

\begin{table}
\caption{\label{tab:table1} The per cent change in area $\Delta A$ and the volume $\Delta V$ of a vesicle during the deformation in a DC electric field. For these data, the computations are done with the same set of parameters are considered, which are used for the figures in the bottom row of Fig.~\ref{fig:shapesCa4p6}.}        
\begin{ruledtabular}
\begin{tabular}{lcr}
$N_{elem}$&$\Delta A$ $(\%)$ \footnote{Note: $\Delta A=|\frac{A-A_0}{A_0}|\times 100$}&$\Delta V$ $(\%)$ \footnote{Note: $\Delta V=|\frac{V-V_0}{V_0}|\times 100$}\\
\hline
-   & ($t=t_m/10,\ t_m/2,\ t_m$)& ($t=t_m/10,\ t_m/2,\ t_m$)\\
\hline 
100 & 0.0200,\ 0.0031,\ 0.0092 & 0.0491,\ 0.0467,\ 0.0524\\
140 & 0.0093,\ 0.0017,\ 0.0067 & 0.0243,\ 0.0287,\ 0.0293\\
180 & 0.0062,\ 0.0009,\ 0.0028 & 0.0139,\ 0.0146,\ 0.0158\\
220 & 0.0041,\ 0.0006,\ 0.0019 & 0.0089,\ 0.0098,\ 0.0101\\
260 & 0.0029,\ 0.0004,\ 0.0013 & 0.0061,\ 0.0070,\ 0.0076\\
300 & 0.0020,\ 0.0002,\ 0.0009 & 0.0046,\ 0.0046,\ 0.0034\\
\end{tabular}
\end{ruledtabular}
\end{table}

Based on the data presented in table~\ref{tab:table1}, $220$ elements and a constant time-step size $\Delta t=10^{-5}$ are considered so that the relative errors in the change in area and volume remain below $0.005\%$ and $0.01\%$, respectively. This set of optimized parameters allow the computation with very high accuracy.

 \section*{DATA AVAILABILITY}
The data that support the findings of this study are available from the corresponding author upon reasonable request.

\bibliography{vesicle_axi_EHD}

\end{document}